\documentclass[%
 reprint,
 amsmath,amssymb,
 aps,
]{revtex4-2}

\usepackage{graphicx}% Include figure files
\usepackage{changes}
\usepackage{dcolumn}% Align table columns on decimal point
\usepackage{bm}% bold math
\usepackage{physics}
\usepackage{float}
\usepackage{placeins}
\usepackage{blindtext}% dummy text
\usepackage{microtype}
\usepackage{xcolor}

\newtheorem{remark}{Remark}

\begin{document}

\preprint{APS/123-QED}

\title{Open harmonic chain without secular approximation}

\author{Melika Babakan$^{1,2}$, Fabio Benatti$^{2,3}$ and Laleh Memarzadeh$^{1}$}
\affiliation{$^1$Physics Department, Sharif University of Technology, Tehran, Iran\\
$^2$Physics Department, University of Trieste, Trieste, Italy\\
$^3$Istituto Nazionale di Fisica Nucleare (INFN), Sezione di Trieste, Trieste, Italy}

\begin{abstract}
We study particle and energy transport in an open quantum system consisting of a three harmonic oscillator chain coupled to thermal baths at different temperatures placed at the ends of the chain. We consider the exact dynamics of the open chain and its so-called local and global Markovian approximations. By comparing them,  we show that, while all three yield a divergence-like continuity equation for the probability flow, the energy flow exhibits instead a distinct behavior. The exact dynamics and the local one preserve a standard divergence form for the energy transport, whereas the global open dynamics, due to the rotating wave approximation (RWA), introduces non-divergence sink/source terms. These terms also affect the continuity equation in the case of a master equation obtained through a time-coarse-graining method whereby RWA is avoided through a time-zoom parameter $\Delta t$. In such a scenario, sink and source contributions are always present for each $\Delta t>0$.  While in the limit $\Delta t\to+\infty$ one recovers the global dissipative dynamics, sink and source terms instead 
vanish  when $\Delta t\to 0$, restoring the divergence structure of the exact dynamics. Our results underscore how the choice of the dissipative Markovian approximation to an open system dynamics critically influences the energy transport descriptions, with implications for discriminating among them and thus, ultimately, for the correct modeling of the time-evolution of open quantum many-body systems.
\end{abstract}
\maketitle
\section{Introduction}

Transport properties of quantum systems are a subject of research in condensed matter physics and  quantum information science, in particular when the system of interest is in interaction with external systems or environments~\cite{davies_model_1978,R_Alicki_1979,Zoubi-2003,Rivas_2010,Migliore_2011,Werlang-Heat-transport-2015,Guimaraes-Nonequilibrium-quantum-chains-2016,Wichterich-heat-transport-2007,rivas_topological_2017,huelga_vibrations_2013,benatti_bath-assisted_2020, benatti_exact_2021, benenti_fundamental_2017,karki_quantum_2020,karki_nonlinear_2019, pavlov_quantum_2021-2}. Recent results cover a wide range of areas with applications in cold atoms \cite{brantut_thermoelectric_2013}, ion traps \cite{barreiro_open-system_2011}, biological networks, and light-harvesting complexes \cite{caruso_highly_2009, killoran_enhancing_2015}. The theoretical analysis of quantum mechanical features such as the sharing of entanglement in many-body systems \cite{ benatti_exact_2021, benatti_environment_2003, memarzadeh_stationary_2011, sierant_dissipative_2022} and its role in the transport properties of materials, in addition to deepening our knowledge in situations where macroscopic descriptions are no longer valid, will lead to the development of many practical applications in the rapidly developing field of quantum technologies. In the following, we will study some of the transport properties of particles and energy in an open harmonic chain consisting of three interacting harmonic oscillators, the first and last of which are coupled to two quantum thermal baths at different temperatures.  

When quantum systems are immersed in an environment whose presence cannot be neglected, their behavior critically depends on how one deals with both the interactions within the open system and between the open system and its environment. The time-evolution of closed quantum systems without interaction with their environment is unitary and thus reversible. In more realistic scenarios, when the coupling to the environment is weak, but not negligible, one can derive a so-called reduced dynamics for the system alone by tracing away the environment. The resulting reduced time-evolution is not reversible anymore and is described by a semigroup of so-called completely positive trace preserving (CPTP) maps. These are the solutions of master equations that are obtained in the so-called secular approximations known as Markovian master equations~\cite{gorini_completely_1976, lindblad_generators_1976}.
The main difficulty in the derivation of Markovian master equations for open quantum systems is the diagonalization of the system Hamiltonian, an operation becoming far from easy for interacting many-body systems. The approach where collective eigenvalues and eigenstates of the Hamiltonian are used is known as global and we will term global dynamics the resulting reduced time-evolution.
The difficulty of diagonalizing the many-body Hamiltonian is avoided by neglecting the interaction when it is sufficiently weaker than the coupling between system and environment: we will refer to the ensuing reduced time-evolution as to the local dynamics. Of course, the  issue here becomes whether important collective effects might be overlooked that are only due to the inter-particles couplings even if weaker than those with the environment.
There are indeed indications in favor of either approach and there is still no general agreement on which one is more reliable~\cite{Zoubi-2003,Rivas_2010,Migliore_2011,levy_local_2014,Werlang-Heat-transport-2015,Guimaraes-Nonequilibrium-quantum-chains-2016,Landi-nonequilibrium-2016,Trushechkin_local-global-2016,Gerardo-local-global-2017,Hofer_local-global-2017,De-Chiara_2018,Cattaneo_2019,Giovannetti-local-global-2020,benatti_bath-assisted_2020,benatti_exact_2021,scali_local_2021,Lutz-local-global-2022,babakan_open_2025}. Logically, to assess this issue, one should either compare theoretical predictions derivable in the two approaches with a suitable experimental evidence or solve the exact dynamics of system and environment together, trace over the environment degrees of freedom and compare the result, which is exact, with those emerging from the local and global approaches.

The importance of choosing the right model for describing the reduced dissipative dynamics of an interacting open quantum many-body system, in particular to analyze its transport properties is evident. In the following, we consider the probability and energy continuity equations for a three-oscillator harmonic chain whose ends are coupled to bosonic thermal baths at different temperatures. We first show that, according to previous observations~\cite{gebauer_current_2004}, the probability flow is governed by the divergence of a dissipative probability current under the exact reduced dynamics and under those obtained by the local and global approaches.
Different is the case of the energy flow: the exact reduced dynamics leads to standard divergence-like continuity equations, as well as for the local  reduced dynamics; however, in the global approach sink/source terms appear that cannot be recast as the divergence of an energy current.

These sink/source contributions to the energy continuity equation in the global approach are shown to emerge because 
of the rotating-wave approximation (RWA). We therefore consider the continuity equation when the master equation for the reduced dynamics is obtained by a time-coarse-graining approach as in~\cite{Schaller-Brandes}. The merit of such a procedure is to avoid recourse to the RWA and to open the way to the inspection of the dissipative dynamics over different time-scales, by means of a zoom parameter $\Delta t$, the coarsest one, $\Delta t\to+\infty$, corresponding to the standard weak-coupling limit dynamics obtained by means of the RWA.

We show that, as in the other two approaches, also in the time-coarse graining one the probability flow is governed by a divergence-like continuity while the energy flow exhibits a distinct behavior consisting, despite the absence of RWA,  in the presence of sink/source terms, namely contributions  that cannot be expressed as divergences  of currents. They are present for all zoom parameters $\Delta t>0$ and vanish only when $\Delta t \to 0$, thereby restoring a divergence-like continuity equation for the energy flow. This result highlights the critical role of approximation choices in shaping the dynamical equations governing energy transport in open many-body quantum systems.

\section{Model}
\label{sec:model}

Denoting by $\rm S$ the harmonic chain and by  $\rm E$ the environment consisting of two thermal baths, one coupled with the first oscillator and the other with the third one of the chain, the total Hamiltonian $H_{\rm T}$ of the compound system and environments is taken of the form 
\begin{equation}
\label{Hamtotal}
\hat{H}_{\rm T}=\hat{H}_{\rm S} +\hat{H}_{\rm E} +\lambda\, \hat{H}',
\end{equation}
where $\lambda\ll1$ is a small dimensionless coupling strength,
\begin{equation}
\label{HS}
\hat{H}_{\rm S}=\omega_0\sum_{i=1}^3a_i^{\dagger}a_i+g\sum_{i=1}^2 \Big(a_i a_{i+1}^{\dagger}+a_i^\dagger a_{i+1}\Big),
\end{equation}
where $a_i$ ($a_i^\dag$), $i=1,2,3$ such that $[a_i,a^\dag_j]=\delta_{ij}$ are the oscillator's annihilation (creation) operators,
$\omega_0>0$ is the common frequency of the three equal Harmonic oscillators and $g>0$ is the strength of their couplings.
Further, $\hat{H}_{\rm E}$ and $\hat{H}'$ are the bath and interaction Hamiltonians, which are taken to be gauge invariant: 
\begin{eqnarray}
\label{HB}
\hat{H}_{\rm E}&=&\sum_{\alpha=\rm{L,R}}\hat{H}_\alpha=\sum_{\alpha=\rm{L,R}}\sum_k\omega_{k\alpha}b^\dag_{k\alpha}b_{k\alpha},
\\
\nonumber
\hat{H}'&=&\sum_{k}\gamma_{k{\rm L}}\Big(a_1^{\dagger}b_{k{\rm L}}+a_1b^\dag_{k{\rm L}}\Big)\\
\label{Hint}
&+&
\sum_{k}\gamma_{k{\rm R}}\Big(a_3^{\dagger}b_{k{\rm R}}+a_3b^\dag_{k{\rm R}}\Big),
\end{eqnarray}
with $b_{j\alpha}$, $b^\dag_{j\alpha}$ the annihilation and creation operators  of  the left (L) and right (R) bath; they satisfy $[b_{j\alpha},b^\dag_{k\beta}]=\delta_{\alpha\beta}\delta_{jk}$ and $[a_j,b_{k\alpha}]=[a_j,b^\dag_{k\alpha}]=[a^\dag_j,b_{k\alpha}]=0$.

\begin{figure}
\centering
\includegraphics[width=0.8\linewidth]{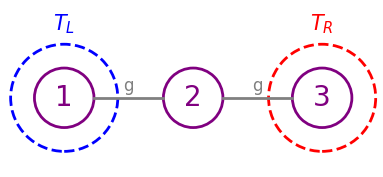}
\caption{The schematic representation of a three interacting harmonic oscillators which also interact with multi-mode external thermal baths at the two ends with temperatures $T_\ell$ and $T_r$.}
\label{fig:schematic representation}
\end{figure}
The exact unitary dynamics of the system and the two baths is given by the standard Liouville-von Neumann equation,
\begin{equation}
\label{eq:von-neumann}
    \frac{d}{dt}\rho_{\rm T}(t)=-i[\hat{H}_{\rm T},\ \rho_{\rm T}(t)],
\end{equation}
where $\rho_{\rm T}(t)$ is the density matrix of the harmonic chain and the two baths.
Due to the infinite number of degrees of freedom of the environment, its states should properly be described as generic expectation functionals on the environment algebra of observable; however, as has become common in the literature on open quantum systems~\cite{breuer_theory_2007}, we will  denote the evolving state of system plus environment with the symbol $\rho_{\rm T}(t)$ used for density matrices.
As customary when dealing with open quantum systems, we shall consider initially compound factorized states $\rho_S(0)\otimes\rho_L\otimes\rho_R$ where $\rho_S(0)$ is a generic state of the harmonic chain whereas $\rho_{R,L}$ are thermal equilibrium Gibbs states at inverse temperatures $\beta_{L,R}$:
\begin{equation}
\label{eq-state}
\rho_{\rm T}(0)=\rho_S(0)\otimes\rho_L\otimes\rho_R,\quad \rho_{\alpha}=\frac{1}{Z_{\alpha}}
{\rm e}^{-\beta_{\alpha}\hat{H}_{\alpha}}.
\end{equation}
Notice that, in line of principle, the exact dynamics of the harmonic chain can always be derived by tracing over the environment degrees of freedom 
and is formally given by the dynamical map:
\begin{equation}
\label{exactdyn}
\rho_S(0)\mapsto\rho_S(t)\equiv{\rm Tr}_{\rm E}\big(\rho_{_{\rm T}}(t)\big),
\end{equation}
where $\operatorname{Tr}_E$ is the trace over the environment Hilbert space.

\subsection{Markovian global and local master equations}
\label{sec:markov-master-equation}

When 1) the coupling of system and environment is sufficiently weak, and 2) there is a clear separation between the time scale of the quantum system without environment  and the 
time-scale of the time-correlation functions of the latter, the so-called weak coupling limit (WCL) is applicable. This limit allows to eliminate the environment in a way that the exact 
evolution in Eq.~(\ref{exactdyn}) is very well approximated by a master equation of the form $\dot{\rho}_S(t)=\mathcal{L}[\rho_S(t)]$ with a generator $\mathcal{L}$ in the  typical Gorini-Kossakowski-Sudarshan-Lindblad (GKSL) form,
thereby ensuring the complete positivity and trace-preserving features of the generated maps:
\begin{equation}
\Phi_t=\exp(t\mathcal{L}):\rho_S(0)\mapsto\rho_S(t),\quad t\geq 0 .
\end{equation}
A short survey of the standard derivation of the generator 
\begin{eqnarray}
\nonumber
    &&\hskip-.6cm\mathcal{L}[\rho_S(t)]=-i[\hat{H}_S+\lambda^2 \hat{H}_{\rm LS},\rho_S(t)]\\
    \label{eq:GKSLME}
    &&\hskip-.7cm
    +\sum_{i,k}(A_i(\omega_k)\rho_S(t)A_i^\dag(\omega_k)-\frac{1}{2}\{A_i^\dag(\omega_k)A_i(\omega_k),\rho_S(t)\}),
\end{eqnarray}
of the GKSL master equation~\cite{gorini_completely_1976,lindblad_generators_1976} is provided in Appendix~\ref{app1} (see in particular Eqs.~(\ref{eq:secular-approximation-Gen})~(\ref{HLS}) and~(\ref{SOp})).
The global and local approaches  to the derivation of the master equation differ exactly in the choice of the eigen-projections of energy $P_{\epsilon}$ in Eq.~(\ref{eq:LindOp}).
While in the global approach they are chosen to be those of $\hat{H}_S$, in the local one they are instead those of the system Hamiltonian $\hat{H}^0_S=\omega_0\sum_{i=1}^3a_i^{\dagger}a_i$ where all interactions among the parties have been switched off. 

\subsubsection{Global master equation}

The diagonalization of the system Hamiltonian $\hat H_S$  in Eq.~(\ref{HS}) is achieved by means of the Bogolubov transformation to new annihilation operators $c_i$ satisfying $[c_i\,,\,c^\dag_j]=\delta_{ij}$ and given by
\begin{equation}
\label{eq:xiglb}
    c_i=\sum_k\mathsf{T}_{ik}a_k,\quad  \mathsf{T}=[\mathsf{T}_{ik}]=\frac{1}{2}\begin{pmatrix}
        1&-\sqrt{2}&1\\
        \sqrt{2}&0&-\sqrt{2}\\
        1&\sqrt{2}&1\\
    \end{pmatrix}.
\end{equation} 
Then, the chain Hamiltonian in Eq.~(\ref{HS}) reads
\begin{equation}
\label{Hd}
    \hat{H}_{\rm S}=\sum_{i=1}^3\epsilon_ic_i^{\dagger}c_i, \quad 
\left\{\begin{matrix}
   \epsilon_1&=&\omega_0-\sqrt{2}g\cr
    \epsilon_2&=&\omega_0\cr
    \epsilon_3&=&\omega_0+\sqrt{2}g
    \end{matrix}\right.,
\end{equation}
with eigen-energies 
\begin{equation}
\label{eigenerg}
E_{n_1,n_2,n_3}=\sum_{i=1}^3 n_i\epsilon_i, \quad n_i\in\mathbb{N} ,
\end{equation}
and eigenvectors
\begin{equation}
\ket{n_1,n_2,n_3}=
\frac{(c^\dagger_1)^{n_1}(c^\dagger_2)^{n_2}(c^\dagger_3)^{n_3}}{\sqrt{n_1!n_2!n_3!}}\vert 0\rangle,
\end{equation}
where $c_i\vert 0\rangle=0$, $i=1,2,3$.

Among all possible transitions, the system-bath interaction in Eq.~(\ref{Hint}) only allows for transition frequencies equal to $\epsilon_i$, for $i=1,2,3$. Due to the coupling with a thermal Bosonic environment with positive energies, the rotating-wave approximation at the root of the weak-coupling limit procedure implies that the dissipative effects of the environment only depend on positive transition frequencies. While $\epsilon_2$ and $\epsilon_3$ are always positive, $\epsilon_1$ is positive only for $g<\frac{\omega_0}{\sqrt{2}}$. This means that for $g\geq\frac{\omega_0}{\sqrt{2}}$ the transition frequency $\epsilon_1$ will not contribute with any Lindblad operator to the GKSL generator of the global dynamics, $\mathcal{L}_{\rm glb}$, which is the form \cite{babakan_open_2025}:
\begin{align}
\label{eq:LGlobal}
    \mathcal{L}_{\rm glb}[\rho_S(t)]&=-i[\hat{H}_S+\lambda^2\hat{H}^{\rm (LS)}_{\rm glb},\rho_S(t)]\cr
    &+2\pi\lambda^2\sum_{i=1}^3\mathcal{D}^{(i)}_{\rm glb}[\rho_S(t)]\ .
\end{align}
The Lamb-Shift correction to the Hamiltonian is given by
\begin{equation}
\label{globalLS}
    \hat{H}^{\rm (LS)}_{ \rm glb}=\sum_iS(\epsilon_i)c_i^\dag c_i, 
\end{equation}
where the real-valued function $S(\epsilon_i)$ reads
\begin{equation}
\label{eq:Sglobal}
S(\epsilon_i)=\sum_{\alpha=L,R}S_{\alpha}(\epsilon_i)=\sum_{\alpha=L,R}\mathcal{P}.\mathcal{V}\int_{0}^{+\infty}
d\omega' \frac{J_{\alpha}(\omega')}{\epsilon_i-\omega'}.
\end{equation}
The dissipative terms in Eq.~(\ref{eq:LGlobal}) are instead given by
\begin{align}
    \mathcal{D}^{(i)}_{\rm glb}[\bullet&]=\frac{1}{4}\sum_{\alpha=L,R}J_{\alpha}(\epsilon_i)\Big(\bar{n}_{\alpha}(\epsilon_i)(c_i^{\dagger}\bullet c_i-\frac{1}{2}\{c_ic_i^{\dagger},\bullet\})\nonumber\\
    &+(\bar{n}_{\alpha}(\epsilon_i)+1)(c_i\bullet c_i^{\dagger}-\frac{1}{2}\{c_i^{\dagger}c_i,\bullet\})\Big),\quad i=1,3,\nonumber\\
    \mathcal{D}^{(2)}_{\rm glb}[\bullet]&=
\frac{1}{2}\sum_{\alpha=L,R}J_{\alpha}(\epsilon_2)\Big(\bar{n}_{\alpha}(\epsilon_2)(c_2^{\dagger}\bullet c_2-\frac{1}{2}\{c_2c_2^{\dagger},\bullet\})
\nonumber\\  
&+(\bar{n}_{\alpha}(\epsilon_2)+1)(c_2\bullet c_2^{\dagger}-\frac{1}{2}\{c_2^{\dagger}c_2,\bullet\})\Big).
\label{globalLind2}
\end{align}
Here, $J_{\alpha}(\epsilon_i)$ are the bath spectral densities, which, for sake of simplicity, we choose to be equal,  $J_L(\epsilon_i)=J_R(\epsilon_i)=J(\epsilon_i) $, and of  Ohmic type with cut-off frequency $\omega_c$:
\begin{equation}
\label{eq:spectral-density}   
J(\epsilon_i):=\sum_k\gamma_k^2\delta(\epsilon_i-\omega_k)=\epsilon_i e^{-\frac{\epsilon_i}{\omega_{c}}} .
\end{equation}
Furthermore,  $\bar{n}_{\alpha}(\epsilon_i)$ denotes the $\alpha$-bath mean photon number at energy $\epsilon_i$ and temperature $T_\alpha$: 
\begin{equation}
\label{eq:mean-photon}
\bar{n}_{\alpha}(\epsilon_i)=\frac{1}{e^{\beta_{\alpha}\epsilon_i}-1},\quad \beta_{\alpha}=\frac{1}{T_{\alpha}},
\end{equation}
where we have set the Boltzmann constant $\kappa_B$ equal to 1.

\subsubsection{Local master equation}

In the local approach, the system Hamiltonian considered in the weak-coupling limit procedure is not the off-diagonal $\hat{H}_S$, rather the already diagonal non-interacting one $\hat{H}^0_S=\omega_0\sum_{i=1}^3 a_i a_i$. Then, the projectors in Eq.~(\ref{eq:LindOp}) are projectors onto the eigenstates of $\hat{H}^0_S$. 
Then, the GKSL generator of the local dynamics, $\mathcal{L}_{\rm loc}$, reads \cite{babakan_open_2025}:
\begin{align}
\label{eq:masterlocal}
   \mathcal{L}_{\rm loc}[\rho_S(t)]&=-i[\hat{H}_S+\lambda^2\hat{H}^{\rm (LS)}_{\rm loc},\rho_S(t)]\cr
   &+2\pi\lambda^2\sum_{\alpha=L,R}\mathcal{D}_{\alpha,\rm loc}[\rho_S(t)].
\end{align}
Up to a negligible multiple of the identity, the Lamb-Shift correction in the local master equation Eq.~(\ref{eq:masterlocal}) is  
\begin{equation}
    \hat{H}^{\rm (LS)}_{ \rm loc}=S_L(\omega_0)a_1^\dag a_1+S_R(\omega_0)a_3^\dag a_3,
\end{equation}
where $S_{\alpha}(\omega_0)$ is a real-valued function:
 \begin{equation}
 \label{eq:principalvalue}
    S_{\alpha}(\omega_0)=\mathcal{P}.\mathcal{V}\int_{0}^{+\infty}d\omega \frac{J_{\alpha}(\omega)}{\omega_0-\omega}.
\end{equation}
Instead, the dissipative contributions to the generator are 
\begin{align}
\mathcal{D}_{L,\rm loc}[\bullet]&=J_{L}(\omega_0)\Big( \bar{n}_L(\omega_0)(a_1^{\dagger}\bullet a_1-\frac{1}{2}\{a_1a_1^{\dagger},\bullet\})\cr
    &+(\bar{n}_{L}(\omega_0)+1)(a_1\bullet a_1^{\dagger}-\frac{1}{2}\{a_1^{\dagger}a_1,\bullet\})\Big) ,\\
\mathcal{D}_{R,\rm loc}[\bullet]&=J_{R}(\omega_0)\Big(\bar{n}_R(\omega_0)(a_3^{\dagger}\bullet a_3-\frac{1}{2}\{a_3a_3^{\dagger},\bullet\})\cr
    &+(\bar{n}_{R}(\omega_0)+1)(a_3\bullet a_3^{\dagger}-\frac{1}{2}\{a_3^{\dagger}a_3,\bullet\})\Big),
\end{align}
where spectral densities and mean-photon numbers are as in the global approach.

\subsection{Time coarse-graining master equation}
\label{sec:coarse-grained-master-equation}

We briefly overview of the time-coarse-graining (TCG) approach to open system dynamics as developed in~\cite{Schaller-Brandes}, more details can be found in Appendix~\ref{CG}.
As already remarked in the Introduction, such an approach  avoids the recourse to the RWA; yet it guarantees the complete positivity of the generated dynamics. Indeed, one again exploits the smallness of the coupling between $\rm S$ and $\rm E$ and the fastness of the environment time-scale compared to the system, but also introduces a time-zoom parameter $\Delta t$ that serves to inspect various time-scales. As discussed in detail in \cite{Benatti-Entanglement-2010}, one focuses upon the variation  of the reduced state 
$\tilde\rho_S(t)\equiv{\rm Tr}_{\rm E}\big(\tilde\rho_{_{\rm T}}(t)\big)$ 
over an externally fixed time-interval $\Delta t$, where $\tilde{\rho}_T(t)$ is the total density matrix in the interaction picture, $\tilde{\rho}_T(t)=e^{it\hat{H}_0}\rho_T(t)e^{-it\hat{H}_0}$ with $\hat{H}_0=\hat{H}_S+\hat{H}_E$.
By taking the trace over the environment 
of the integrated version of Eq.~(\ref{7}) and by truncating the Dyson expansion at the lowest order
in $\lambda$, one gets:
\begin{eqnarray}
\nonumber
\hskip-.5cm
&&
\frac{\tilde\rho_S(t+\Delta t)-\tilde\rho_S(t)}{\Delta t}\\
\hskip-.5cm
&&
=-\frac{\lambda^2}{\Delta t}\int_t^{t+\Delta t} \hskip-.7cm dt_1
\int_t^{t_1} \hskip-.4cm dt_2{\rm Tr}_{\rm E}\Big( \big[ \tilde{H}'(t_1),\big[\tilde{H}'(t_2),\, \tilde\rho_{\rm T}(t)\big]\big]\Big),
\label{8}
\end{eqnarray}
where
\begin{equation}
    \tilde{H}'(s)=e^{is\hat{H}_0}\ \hat{H}'\ e^{-is\hat{H}_0}.
\end{equation}
On the other hand, the l.h.s of Eq.~(\ref{8}) is
\begin{equation}
\label{eq:derivative}
    \frac{\tilde\rho_S(t+\Delta t)-\tilde\rho_S(t)}{\Delta t}=
\frac{1}{\Delta t}\int_t^{t+\Delta t} ds\ \frac{\partial\tilde\rho_S(s)}{\partial s},
\end{equation}
by using the chain derivative and changing the variable $\tau=\lambda^2t$ the r.h.s of Eq.~(\ref{eq:derivative}) becomes
\begin{equation}
{\tilde\rho_S(t+\Delta t)-\tilde\rho_S(t)\over \Delta t}=
{1\over\Delta t}\int_\tau^{\tau+\lambda^2\Delta t} ds\ {\partial\tilde\rho_S(s/\lambda^2)\over \partial s}.
\label{9}
\end{equation}
Therefore, in the limit of $\lambda^2\Delta t\to 0$ for small $\lambda$ and finite $\Delta t$,
one can readily approximate the r.h.s. of Eq.~(\ref{9}) with $\partial_t\tilde\rho_S(t)$.
At this point, one further observes that the environment is in general so much larger than the subsystem immersed in it that its
dynamics is hardly affected by its presence. Therefore, it is justified to replace 
in the double integral of Eq.~(\ref{8}) the evolved total state $\tilde\rho_{\rm T}(t)$ with
the product state $\tilde\rho_S(t)\otimes\rho_{\rm E}$ with $\rho_{\rm E}=\rho_L\otimes\rho_R$, taking the initial environment state in Eq.~(\ref{eq-state}) as bath reference state~\cite{gorini_completely_1976}-\cite{alicki_quantum_1987}.
Returning to the Schr\"odinger representation, one finally gets the following time-coarse-grained Markovian master equation for the system \cite{Schaller-Brandes}:
\begin{align}
\label{10}
\frac{d}{dt}\rho_S(t)&= -i \big[\hat{H}_S+\lambda^2 \hat{H}^{(\rm LS)}_{\rm tcg}(\Delta t),\, \rho_S(t)\big]
 + {\cal D}_{\rm tcg}[\rho(t)]\cr
 &:={\cal L}_{\rm tcg}[\rho_S(t)],
\end{align}
where the Lamb-shift Hamiltonian correction is
\begin{align}
\label{eq:HLambTCG}
  \hat{H}^{({\rm LS})}_{\rm tcg}(\Delta t)&={i\lambda^2\over2\Delta t}\int_0^{\Delta t}ds_1\int_0^{\Delta t}
\theta(s_1-s_2)\cr
&\times {\rm Tr}_E\Big(\rho_E\big[ \tilde{H}'(s_1),\tilde{H}'(s_2)\big]\Big),
\end{align}
while the dissipation part is
\begin{align}
  {\cal D}_{\rm tcg}[\rho_S(t)]&={\lambda^2\over \Delta t} {\rm Tr}_E\Big(
L(\Delta t)\,\big(\rho_S(t)\otimes\rho_E\big)\, L(\Delta t)\cr
&-{1\over2}\Big\{ L^2(\Delta t),\big(\rho_S(t)\otimes\rho_E\big)\Big\}\Big),
\label{13}  
\end{align}
with 
\begin{equation}
  L(\Delta t)=\int_0^{\Delta t}ds\, \tilde{H}'(s).  
\end{equation}
The curly brackets representing the anti-commutator, while $\theta(s)$ is the step function.
It is important to observe that, for any interval $\Delta t$,  Eqs.~(\ref{10})-(\ref{13}) generate
a quantum dynamical semigroup of completely
positive maps. Indeed, the dissipation in the r.h.s. of Eq.~(\ref{13}) turns out to be itself completely positive,
the composition of two completely positive maps, the trace over the environment degrees
of freedom and a linear operator on the total system, being all written in the canonical Kraus-Stinespring form \cite{takesaki_theory_1979,alicki_quantum_2001}.
Notice that, on the contrary, in the weak-coupling limit complete positivity is ensured by an ergodic average prescription, that, as mentioned in the
Introduction, eliminates fast oscillating terms (RWA)~\cite{dumcke_proper_1979,davies_markovian_1974,davies_markovian_1976}; 
in the present formalism, this corresponds
to letting the time-coarse-graining parameter $\Delta t$ going to infinity: it is thus applicable
only to environments with sharp decaying correlations. 

As derived in detail in Appendix~\ref{CG}, the time-coarse-graining $\Delta t$-dependent GKSL generator $\mathcal{L}_{tcg}$ reads:
\begin{align}
\label{eq:CG bosonic chain}
    \mathcal{L}_{tcg}[\rho_S(t)]&=-i[\hat{H}_{\rm S}+\lambda^2\hat{H}_{\rm tcg}^{(\rm LS)}(\Delta t),\rho_S(t)]\cr
    &+\lambda^2\sum_{i,j=1}^3\mathcal{D}^{(ij)}_{\rm tcg}[\rho_S(t)].
\end{align}
The Lamb-shift Hamiltonian, $H^{(\rm LS)}_{\rm tcg}(\Delta t)$ is
\begin{equation}
\label{eq:lamb3harmonic}
    H^{(\rm LS)}_{\rm tcg}(\Delta t)=\sum_{i,j=1}^3(S^-_{ij}(\Delta t)c_ic_j^\dag+S^+_{ij}(\Delta t)c_i^\dag c_j)\ , 
\end{equation}
with 
\begin{align}
\label{lambccoeff}
    &S^{\pm}_{ij}(\Delta t)=\frac{2\Delta t}{i\pi}e^{-i(\epsilon_i-\epsilon_j)\frac{\Delta t}{2}}\int_{-\omega_c}^{\omega_c}d\omega\frac{f^{\pm}(\omega)}{(\omega-\epsilon_j)}\times \cr
    &\hskip-.2cm\Big(\operatorname{sinc}\big ((\epsilon_i-\epsilon_j)\frac{\Delta t}{2}\big) \hskip-.1cm-\hskip-.1cm\operatorname{sinc}{\big((\omega-\epsilon_i)\frac{\Delta t}{2}\big)}\operatorname{cos}{\big((\omega-\epsilon_j)\frac{\Delta t}{2}\big)}\Big),\cr
\end{align}
and the function $\displaystyle\operatorname{sinc}(x)=\frac{\sin x}{x}$. Indeed, the dissipative part of the time-coarse-graning master equation is
\begin{align}
\label{eq:disTCG}
    \mathcal{D}^{(ij)}_{\rm tcg}[\bullet]&=\gamma^+_{i,j}(\Delta t)\left(c_i\bullet c_j^\dag -\frac{1}{2}\{c_j^\dag c_i,\bullet\}\right)\cr
    &+\gamma^-_{i,j}(\Delta t)\left(c_i^\dag\bullet c_j -\frac{1}{2}\{c_j c_i^\dag,\bullet\}\right).
\end{align}
where the coefficients, $\gamma^{\pm}_{i,j}(\Delta t)$ in Eq.~(\ref{eq:disTCG}) are
\begin{align}
\label{eq:gamma}
    \gamma_{ij}^{\pm}(\Delta t)&=\frac{\Delta  t}{8\pi}e^{i(\epsilon_j-\epsilon_i)\frac{\Delta t}{2}}\int_{-\omega_c}^{\omega_c}d\omega f^{\pm}_{ij}(\omega) \cr
    &\times\operatorname{sinc}\big((\omega-\epsilon_i)\frac{\Delta t}{2}\big)\operatorname{sinc}\big((\omega-\epsilon_j)\frac{\Delta t}{2}\big),
\end{align}
with
\begin{equation}
\label{eq:ffunc}
    f_{ij}^{\pm}(\omega) =
    \begin{cases}
        \tau^{\pm}(\omega)(1 + \delta_{i2}), & i = j, \\
        \sqrt{2}(J_R(\omega)\bar{n}_R(\omega) - J_L(\omega)\bar{n}_L(\omega)), & |i-j| = 1, \\
        \tau^{\pm}(\omega), & |i-j| = 2,
    \end{cases}
\end{equation}
where 
\begin{align}
\label{eq:tauplus}
    &\tau^+(\omega)=\sum_{i=1,2}\sigma_{ii}(\omega)=\sum_{\alpha=L,R}J_{\alpha}(\omega)(\Bar{n}_{\alpha}(\omega)+1),\\
    \label{eq:tauminus}
    &\tau^-(\omega)=\sum_{i=3,4}\sigma_{ii}(\omega)=\sum_{\alpha=L,R}J_{\alpha}(\omega)\Bar{n}_{\alpha}(\omega).
\end{align}
In the above expressions, the quantities $\sigma_{k\ell}(\omega)$ are the Fourier transforms of the environment time-correlation functions ${\rm Tr}_E\Big(\rho_E B_k(s)B_\ell\Big)$.
Namely,
\begin{equation}
\label{correlations2}
\sigma_{k\ell}(\omega):=\frac{1}{2\pi}\int_{-\infty}^{+\infty}ds\,e^{-i\omega s}{\rm Tr}_E\Big(\rho_EB_k(s)B_\ell\Big),
\end{equation}
with $B_1$ and $B_3$, respectively $B_2$ and $B_4$ operators of the left, respectively right bath as in Eq.~(\ref{opB}) in Appendix~\ref{CG}.
Then, because of the assumed thermal structure  of the environment state,
$\rho_E=\rho_L\otimes\rho_R$, only contributions with $k=\ell$ remain.
Further, in Eqs.~(\ref{eq:ffunc}--\ref{eq:tauminus}), $\Bar{n}_{\alpha}(\omega)$ is the mean photon number in each bath, which has been defined in Eq.~(\ref{eq:mean-photon}),
while $J_{\alpha}(\omega)$ is the spectral density of bath $\alpha=L,R$, which has been defined in Eq.~(\ref{eq:spectral-density}).
As shown in Appendix~\ref{App:2}, the standard weak-coupling limit result exactly corresponds to the coarsest grain, $\Delta t\to+\infty$.

\begin{remark}
\label{rem:Gauss}
Being at most quadratic in creation and annihilation operators, the generators $\mathcal{L}$ in the various approaches are such that Gaussian states are mapped into Gaussian states~\cite{OlivaresParis}. Gaussian states are completely determined by their covariance matrices~\cite{serafini_quantum_2021}; the consequences of this fact are two-fold (see Appendix~\ref{app:CovTcg}): on one hand, the open reduced dynamics of Gaussian states can be recast in terms of a time-evolution equation of their covariance matrices. On the other hand, because of the structure of the generator, only the identity operator commutes with all the contributing  Lindblad operators, Hamiltonian included. Therefore, there exist a unique stationary state such that $\partial_t\rho_t=\mathcal{L}[\rho_t]=0$ and this latter must then be Gaussian and thus specified by a unique asymptotic  covariance matrix. Such properties will become of particular use later on in what follows when we shall deal with the asymptotic transport features.
\end{remark}

\section{Probability density continuity equation}

Transport features of many-body quantum systems  are theoretically assessed by means of suitable continuity equations; these latter may involve diverse quantities as  probability, particle-numbers, energy, angular momentum and spin. It is of noticeable importance to inspect how  the presence of dissipation, noise and decoherence affects them.
The standard continuity equation describes how the probability density $\rho_t(\boldsymbol{x})$, $\boldsymbol{x}\in\mathbb{R}^3$, 
associated with a quantum state $\rho(t)$ varies in time.
For a unitary dynamics generated by a Hamiltonian with a local potential, such a variation is due only to the divergence of the probability current $\boldsymbol{j}_t(x)$ without sinks, respectively sources absorbing, respectively creating probability; namely,
\begin{equation}
\label{eq:continuity}
    \partial_t\rho_t(\boldsymbol{x})+\nabla \,\boldsymbol{j}_t(\boldsymbol{x})=0.
\end{equation}
Together, these terms ensure that any increase, respectively decrease of the probability density within a finite volume in space is due to an incoming, respectively outgoing flow of probability through its surface.
For $N$ quantum particles moving in $\mathbb{R}^3$, the probability density at the spatial position $\boldsymbol{x}\in\mathbb{R}^3$ and time $t$ is
\begin{equation}
  \rho_t(\boldsymbol{x})={\rm Tr}\big(\hat{n}(\boldsymbol{x})\rho(t)\big),  
\end{equation}
where the spatial density operator, $\hat{n}(\boldsymbol{x})$, counts the fraction  of particles at the spatial position $\boldsymbol{x}$; namely,
\begin{equation}
\label{op-density}
    \hat{n}(\boldsymbol{x})=\frac{1}{N}\sum_{i=1}^N\delta(\boldsymbol{x}-\hat{\boldsymbol{x}}_i),
\end{equation}
with $\hat{\boldsymbol{x}}_i$ the $i$-th particle $3$-dimensional position operator.
If the particles interact through a spatially local potential of the form
$\displaystyle \hat{H}_S=\sum_{j=1}^N\frac{\hat{\boldsymbol{p}}_j^2}{2}+V(\hat{\boldsymbol{x}}_j)$,
then (see~\cite{PhysRevLett.93.160404},~\cite{PhysRevA.73.064101}), the operator current associated with the operator density in Eq.~(\ref{op-density}), $\widehat{\boldsymbol{j}}(\boldsymbol{x})$, is computed as 
\begin{equation}
\label{op-current}
    \widehat{\boldsymbol{j}}(\boldsymbol{x})=\frac{1}{2N}\sum_{i=1}^N \big(\delta(\boldsymbol{x}-\hat{\boldsymbol{x}}_i)\hat{\boldsymbol{p}}_i + \hat{\boldsymbol{p}}_i\delta(\boldsymbol{x}-\hat{\boldsymbol{x}}_i)\big),
\end{equation}
where $\hat{\boldsymbol{p}}_i$ denotes the $3$-dimensional momentum operator of the $i$-th particle.
Then, the probability current relative to the density matrix $\rho(t)$ is given by
\begin{equation}
\label{prob-current}
 \boldsymbol{j}_t(\boldsymbol{x})={\rm Tr}\big(\widehat{\boldsymbol{j}}(\boldsymbol{x})\rho(t)\big).
\end{equation}
In the  case of the harmonic chain considered here, by going from annihilation and creation operators 
to one-dimensional position and momentum operators for the three oscillators,
\begin{equation}
\label{pos-mom}
a_i=\sqrt{\frac{\omega_0}{2}}\hat{x}_i+\frac{i}{\sqrt{2\omega_0}}\,\hat{p}_i,\quad 
a_i^\dag=\sqrt{\frac{\omega_0}{2}}\hat{x}_i-\frac{i}{\sqrt{2\omega_0}}\,\hat{p}_i,
\end{equation}
the Hamiltonian in Eq.~(\ref{HS}) can be recast, neglecting a scalar term,  as
\begin{equation}
\label{eq:HSxp}
\hat{H}_S=\frac{1}{2}\sum_{i=1}^3
\big(\hat{p}_i^2+\omega_0^2\hat{x}_i^2\big)+\frac{g}{\omega_0}\sum_{i=1}^2\big(\hat{p}_i\hat{p}_{i+1}+\omega_0^2\hat{x}_i\hat{x}_{i+1}\big).
\end{equation}
As shown in  Appendix~\ref{app:Continuity equation}, in such a case the corresponding operator current density $\widehat{j}^{\mathcal{U}}(x)$ is 
\begin{eqnarray}
\label{unppcurr1}
\hskip-.7cm
\nonumber
\widehat j^{\mathcal{U}}(x)&=&\frac{1}{6}\sum_{\ell=1}^3\Big(\delta(x-\hat x_\ell)\hat p_\ell+\hat p_\ell\delta(x-\hat x_\ell)\Big)\\
&+&\frac{g}{3\omega_0}\sum_{\ell=1}^2\Big(\hat p_\ell\delta(x-\hat x_{\ell+1})+\hat p_{\ell+1}\delta(x-\hat x_\ell)\Big).
\label{unppcurr2}
\end{eqnarray}
In~\cite{PhysRevLett.93.160404}, it was proved that, contrary to many expectations,  $N$ electrons interacting locally among themselves and coupled to a thermal phonon bath, via local electron operators, undergo a reduced dissipative dynamics that still supports a continuity equation as in Eq.~(\ref{prob-current}), the presence of the common environment contributing with an extra dissipative current $\boldsymbol{j}^\mathcal{D}_t(\boldsymbol{x})$: 
\begin{equation}
   \partial_t\rho_t(\boldsymbol{x})+\nabla \,\boldsymbol{j}^\mathcal{U}_t(\boldsymbol{x})+\nabla \,\boldsymbol{j}^\mathcal{D}_t(\boldsymbol{x})=0.
 \end{equation}
In the following sub-sections, we study how the probability continuity equation changes for the open harmonic chain without approximating the reduced dynamics and for the global, local and time-coarse-graining approximations.

Given a dissipative dynamics generated by a GKSL generator $\mathcal{L}$, we shall then focus upon 
\begin{eqnarray}
\label{eq:dissconteq1a}
\nonumber
    \partial_t\rho_t(x)&=&\partial_t{\rm Tr}\big(\rho(t)\hat{n}(x)\big)={\rm Tr}\big(\mathcal{L}[\rho(t)]\hat n(x)\big)\\
    \label{eq:dissconteq1b}
    &=&{\rm Tr}\big(\rho(t)\,\mathcal{L}^{\rm adj}[\hat n(x)]\big), 
\end{eqnarray}
where ${\cal L}^{adj}$ denotes the adjoint of the GKSL generator defined by moving from the Schrödinger to the Heisenberg picture. The cyclicity of the trace, ${\rm Tr}(A\,B)={\rm Tr}(B\,A)$, can be used to show that a GKSL generator in Eq.~(\ref{eq:GKSLME})
has its adjoint on operators $X$ of the form
\begin{eqnarray}
&&\nonumber
\hskip-.5cm\mathcal{L}^{\rm adj}[X]=i[\hat{H}_S+\lambda^2 \hat{H}_{\rm LS},X]\\
&&\hskip-.5cm\label{adj_aux}
+\sum_{i,k}(A_i^\dag(\omega_k)XA_i(\omega_k)-\frac{1}{2}\{A_i^\dag(\omega_k)A_i(\omega_k),X\})\ .
\end{eqnarray}

\subsection{Exact reduced dynamics}

We begin by considering the probability continuity equation in the case of the open harmonic chain whose dynamics is generated by Eq.~(\ref{eq:von-neumann}).
The time derivative of the probability density by using the adjoint of the generator in Eq.~(\ref{eq:von-neumann}) ($
\mathcal{L}^{\rm adj}[X]=i[\hat{H}_T,X]
$)reads:
\begin{equation}
\label{aux1}
\partial_t\rho_t(x)=i\,{\rm Tr}\Big(\rho_T(t)\,\Big[\hat{H}_S+\lambda \hat{H}'\,,\,\hat{n}(x)\Big]\,\Big).
\end{equation}
The contribution of the system Hamiltonian $\hat{H}_S$ to the commutator amounts to the divergence of a current 
$j^\mathcal{U}_t(x)=\operatorname{Tr}(\widehat{j}^{\mathcal{U}}(x)\rho(t))$ as in Eq.~(\ref{unppcurr1}).
From Eq.~(\ref{Hint}), the remaining commutator reads
\begin{equation}
\label{eq:dissExc}
    i\lambda \Big[\hat{H}'\,,\,\hat{n}(x)\Big]=-\partial_x\widehat{j}^\mathcal{D}(x),
\end{equation}
where, following the analysis performed in Appendix~\ref{app:Continuity equation}, 
\begin{align}
    \widehat{j}^\mathcal{D}(x)&=\frac{\lambda}{3\sqrt{2\omega_0}}\delta(x-\hat{x}_1)\sum_k\gamma_{kL}\big(b_{kL}-b^\dag_{kL}\big)\cr
    &+\frac{\lambda}{3\sqrt{2\omega_0}}\delta(x-\hat{x}_3)\sum_k\gamma_{kR}\big(b_{kR}-b^\dag_{kR}\big).
\end{align}
Therefore, the continuity equation of the probability density for this system reads
\begin{equation}
   \partial_t\rho_t(x)+\partial_x\,j^\mathcal{U}_t(x)+\partial_x\,j^\mathcal{D}_t(x)=0,
 \end{equation}
with dissipative current $j^\mathcal{D}_t(x)=\operatorname{Tr}\Big(\rho_T(t) \widehat{j}^\mathcal{D}(x)\Big)$.

\subsection{Local approach}
In the case of local approach for the open harmonic chain whose dissipative dynamics is generated by Eq.~(\ref{eq:masterlocal}), the probability continuity equation is 
\begin{equation}
\label{eq:dissconteqloc}
    \partial_t\rho_t(x)+\partial_x\Big(j^\mathcal{U}_t(x)+j^{\mathcal{D}_{\rm loc}}_t(x) +\mathcal{Q}^{\rm loc}_t(x) + \partial_x\, \mathcal{P}^{\rm loc}_t(x)\Big)=0,
\end{equation}
where $j^\mathcal{U}_t(x)=\operatorname{Tr}(\widehat{j}^{\mathcal{U}}(x)\rho(t))$ is the current with $\widehat{j}^{\mathcal{U}}(x)$ as in Eq.~(\ref{unppcurr1}), which is due to the unitary part of the evolution, while
$j^{\mathcal{D}_{\rm loc}}_t(x)=\operatorname{Tr}(\widehat{j}^{\mathcal{D}_{\rm loc}}(x)\rho(t))$ is a dissipative current with 
\begin{align}
\label{eq:jDloc}
&\widehat{j}^{\mathcal{D}_{\rm loc}}(x)=\frac{1}{2}\sum_{\ell=1}^3S_{\ell\ell}^{(\rm loc)}\big(\hat{p}_\ell\delta(x-\hat{x}_\ell)+\delta(x-\hat{x}_\ell)\hat{p}_\ell\big)\cr
&+\sum_{\ell=1}^2\big(\Re(S_{\ell,\ell+1}^{(\rm loc)})-i\Im(\alpha^{+(\rm loc)}_{\ell+1,\ell}-\alpha^{-(\rm loc)}_{\ell,\ell+1})\big)\cr
&\hskip+1cm\times\big(\hat{p}_{\ell+1}\delta(x-\hat{x}_\ell)+\hat{p}_{\ell}\delta(x-\hat{x}_{\ell+1})\big).
\end{align}
As they also appear inside the spatial derivative $\partial_x$, the other two terms $\mathcal{Q}^{\rm loc}_t(x)=\operatorname{Tr}(\widehat{\mathcal{Q}}^{\rm loc}(x)\rho(t))$ and $\partial_x\mathcal{P}^{\rm loc}_t(x)=\partial_x\operatorname{Tr}(\widehat{\mathcal{P}}^{\rm loc}(x)\rho(t))$ give 
dissipative current contributions. They involve the following operators: 
\begin{align}
\label{eq:Qloc}
   &\hskip-0.2cm\widehat{\mathcal{Q}}^{\rm loc}(x)=\sum_{\ell=1}^3(\alpha^{+(\rm loc)}_{\ell\ell}-\alpha^{-(\rm loc)}_{\ell\ell})\hat{x}_\ell\delta(x-\hat{x}_\ell)\cr
   &\hskip-0.2cm-\sum_{\ell=1}^2\big(i\Im(S^{(\rm loc)}_{\ell,\ell+1})-\Re(\alpha^{+(\rm loc)}_{\ell+1,\ell}-\alpha^{-(\rm loc)}_{\ell, \ell+1})\big)\cr
   &\hskip+0.9cm\times\big(\hat{x}_{\ell+1}\delta(x-\hat{x}_\ell)+\hat{x}_{\ell}\delta(x-\hat{x}_{\ell+1})\big),\\
\label{eq:Ploc}
&\hskip-0.2cm \widehat{\mathcal{P}}^{\rm loc}(x)=\frac{1}{4}\sum_{\ell=1}^3(\alpha_{\ell\ell}^{+(\rm loc)}+\alpha_{\ell\ell}^{-(\rm loc)})\delta(x-\hat{x}_\ell),
\end{align}
where 
 \begin{align}
 \label{eq:Slambloc}
     &\hskip -.2cm
     S_{\ell k}^{(\rm loc)}=\lambda^2\operatorname{diag}(S_L(\omega_0),0,S_R(\omega_0))\, ,\\
     \label{eq:discoeffloc}
     &\hskip -.2cm
     \alpha^{+(\rm loc)}=\frac{\lambda^2}{2}\cr
     &\hskip -.2cm
     \operatorname{diag}(J_L(\omega_0)(\bar{n}_L(\omega_0)+1),0,J_R(\omega_0)(\bar{n}_R(\omega_0)+1))\, ,\\
     &\hskip -.2cm
     \alpha^{-(\rm loc)}=\frac{\lambda^2}{2}\operatorname{diag}(J_L(\omega_0)\bar{n}_L(\omega_0),0,J_R(\omega_0)\bar{n}_R(\omega_0)).    
 \end{align}

\subsection{Global approach}

In the case of global approach for the open harmonic chain whose dissipative dynamics is generated by Eq.~(\ref{eq:LGlobal}), the probability continuity equation  is 
\begin{equation}
\label{eq:dissconteqglb}
    \partial_t\rho_t(x)+\partial_x\Big(j^\mathcal{U}_t(x)+j^{\mathcal{D}_{\rm glb}}_t(x) +\mathcal{Q}^{\rm glb}_t(x) + \partial_x\, \mathcal{P}^{\rm glb}_t(x)\Big)=0 ,
\end{equation}
where $j^\mathcal{U}_t(x)=\operatorname{Tr}(\widehat{j}^{\mathcal{U}}(x)\rho(t))$ is a current with $\widehat{j}^{\mathcal{U}}(x)$ as in Eq.~(\ref{unppcurr1}), while
$j^{\mathcal{D}_{\rm glb}}_t(x)=\operatorname{Tr}(\widehat{j}^{\mathcal{D}_{\rm glb}}(x)\rho(t))$ is a dissipative current with 
\begin{align}
\label{eq:jDglb}
&\widehat{j}^{\mathcal{D}_{\rm glb}}(x)=\frac{1}{2}\sum_{\ell=1}^3S_{\ell\ell}^{(\rm glb)}\big(\hat{p}_\ell\delta(x-\hat{x}_\ell)+\delta(x-\hat{x}_\ell)\hat{p}_\ell\big)\cr
&+\sum_{\ell=1}^2\big(\Re(S_{\ell,\ell+1}^{(\rm glb)})-i\Im(\alpha^{+(\rm glb)}_{\ell+1,\ell}-\alpha^{-(\rm glb)}_{\ell,\ell+1})\big)\cr
&\hskip+1cm\times\big(\hat{p}_{\ell+1}\delta(x-\hat{x}_\ell)+\hat{p}_{\ell}\delta(x-\hat{x}_{\ell+1})\big).
\end{align}
Furthermore, the other two dissipative currents, $\mathcal{Q}^{\rm glb}_t(x)=\operatorname{Tr}(\widehat{\mathcal{Q}}^{\rm glb}(x)\rho(t))$ and $\mathcal{P}^{\rm glb}_t(x)=\operatorname{Tr}(\widehat{\mathcal{P}}^{\rm glb}(x)\rho(t))$ involve the operators 
\begin{align}
\label{eq:Qglb}
   &\hskip-0.2cm\widehat{\mathcal{Q}}^{\rm glb}(x)=\sum_{\ell=1}^3(\alpha^{+(\rm glb)}_{\ell\ell}-\alpha^{-(\rm glb)}_{\ell\ell})\hat{x}_\ell\delta(x-\hat{x}_\ell)\cr
   &\hskip-0.2cm-\sum_{\ell=1}^2\big(i\Im(S^{(\rm glb)}_{\ell,\ell+1})-\Re(\alpha^{+(\rm glb)}_{\ell+1,\ell}-\alpha^{-(\rm glb)}_{\ell, \ell+1})\big)\cr
   &\hskip+0.9cm\times\big(\hat{x}_{\ell+1}\delta(x-\hat{x}_\ell)+\hat{x}_{\ell}\delta(x-\hat{x}_{\ell+1})\big),\\
\label{eq:Pglb}
&\hskip-0.2cm \widehat{\mathcal{P}}^{\rm glb}(x)=\frac{1}{4}\sum_{\ell=1}^3(\alpha_{\ell\ell}^{+(\rm glb)}+\alpha_{\ell\ell}^{-(\rm glb)})\delta(x-\hat{x}_\ell),
\end{align}
with 
 \begin{align}
 \label{eq:Slambglb}
     &S_{\ell k}^{(\rm glb)}=\lambda^2(\mathsf{T}^\dag S(\epsilon) \mathsf{T})_{\ell k}\, ,\\
     \label{eq:discoeffglb}
     &\alpha^{\pm(\rm glb)}_{\ell k}=\frac{\lambda^2}{2}(\mathsf{T}^\dag\gamma^{\pm(\rm glb)}\mathsf{T})_{\ell k} \ ,
 \end{align}
where  $S^\pm(\epsilon)=[S^\pm (\epsilon)_{ij}]$ and $\gamma^{\pm(\rm glb)}$ is the matrix with entries the coefficients of
\begin{align}
    \gamma^{+(\rm glb)}&=\operatorname{diag}(\tau^+(\epsilon_1),\tau^+(\epsilon_2),\tau^+(\epsilon_3)),\\
    \gamma^{-(\rm glb)}&=\operatorname{diag}(\tau^-(\epsilon_1),\tau^-(\epsilon_2),\tau^-(\epsilon_3)).
\end{align}
These four terms gives rise to an effective, dissipative current such that the time-variation of the probability density is given by the divergence of a current.

\subsection{Time-coarse-graining approach}

We now show that a dissipative current also appears in the probability continuity equation of the open harmonic chain whose dissipative dynamics is generated by Eq.~(\ref{eq:CG bosonic chain}):
\begin{align}
\label{eq:adjointGen}
    &\mathcal{L}_{tcg}^{\rm adj}[\bullet]=i[\hat{H}_{\rm S}+\lambda^2\hat{H}_{\rm LS}^{({\rm tcg})}(\Delta t)\,,\,\bullet]\cr
    &+\lambda^2\sum_{i,j=1}^3\gamma^+_{i,j}(\Delta t)\Big(c_j^\dag\,\bullet\,c_i -\frac{1}{2}\Big\{c_j^\dag c_i\,,\,\bullet\Big\}\Big)\cr
    &+\lambda^2\sum_{i,j=1}^3\gamma^-_{i,j}(\Delta t)\Big(c_j\,\bullet\, c_i^\dag -\frac{1}{2}\Big\{c_j c_i^\dag\,,\,\bullet\Big\}\Big)\ .
\end{align}
The continuity equation  indeed reads
\begin{equation}
\label{eq:dissconteq2}
    \partial_t\rho_t(x)+\partial_x\Big(j^\mathcal{U}_t(x)+j^{\mathcal{D}_{\rm tcg}}_t(x) +\mathcal{Q}^{\rm tcg}_t(x) + \partial_x\, \mathcal{P}^{\rm tcg}_t(x)\Big)=0,
\end{equation}
where $j^\mathcal{U}_t(x)=\operatorname{Tr}(\widehat{j}^{\mathcal{U}}(x)\rho(t))$ is the current with $\widehat{j}^{\mathcal{U}}(x)$ as in Eq.~(\ref{unppcurr1}), while
$j^{\mathcal{D}_{\rm tcg}}_t(x)=\operatorname{Tr}(\widehat{j}^{\mathcal{D}_{\rm tcg}}(x)\rho(t))$ is a dissipative current with 
\begin{eqnarray}
\nonumber
&&\hskip-0.4cm\widehat{j}^{\mathcal{D}_{\rm tcg}}(x)=\frac{1}{2}\sum_{\ell=1}^3S^{(\rm tcg)}_{\ell\ell}(\Delta t)\big(\hat{p}_\ell\delta(x-\hat{x}_\ell)+\delta(x-\hat{x}_\ell)\hat{p}_\ell\big)\\
\nonumber
&&\hskip-0.4cm+\sum_{\ell=1}^2\big(\Re(S_{\ell,\ell+1}^{(\rm tcg)}(\Delta t))-i\Im(\alpha_{\ell+1,\ell}^{+(\rm tcg)}(\Delta t)-\alpha_{\ell,\ell+1}^{-(\rm tcg)}(\Delta t))\big)\\
&&
\hskip+1cm\times\big(\hat{p}_{\ell+1}\delta(x-\hat{x}_\ell)+\hat{p}_{\ell}\delta(x-\hat{x}_{\ell+1})\big).
\label{eq:jDCG}
\end{eqnarray}
The other contributions $\mathcal{Q}^{\rm tcg}_t(x)=\operatorname{Tr}(\widehat{\mathcal{Q}}^{\rm tcg}(x)\rho(t))$ and $\mathcal{P}^{\rm tcg}_t(x)=\operatorname{Tr}(\widehat{\mathcal{P}}^{\rm tcg}(x)\rho(t))$ involve the following operators: 
\begin{eqnarray}
\nonumber
&&\hskip-0.4cm\widehat{\mathcal{Q}}^{\rm tcg}(x)=\sum_{\ell=1}^3(\alpha_{\ell\ell}^{+(\rm tcg)}(\Delta t)-\alpha_{\ell\ell}^{-(\rm tcg)}(\Delta t))\hat{x}_\ell\delta(x-\hat{x}_\ell)\\
\nonumber
&&\hskip-0.4cm-\sum_{\ell=1}^2\big(i\Im(S_{\ell,\ell+1}^{(\rm tcg)}(\Delta t))-\Re(\alpha_{\ell+1,\ell}^{+(\rm tcg)}(\Delta t)-\alpha_{\ell, \ell+1}^{-(\rm tcg)}(\Delta t))\big)\\
\label{eq:QCG}
&&
\hskip+0.9cm\times\big(\hat{x}_{\ell+1}\delta(x-\hat{x}_\ell)+\hat{x}_{\ell}\delta(x-\hat{x}_{\ell+1})\big),\\
\label{eq:PCG}
&&\hskip-0.5cm
\widehat{\mathcal{P}}^{\rm tcg}(x)=\frac{1}{4}\sum_{\ell=1}^3(\alpha_{\ell\ell}^{+(\rm tcg)}(\Delta t)+\alpha_{\ell\ell}^{-(\rm tcg)}(\Delta t))\delta(x-\hat{x}_\ell),
\end{eqnarray}
with 
 \begin{align}
 \label{eq:SlambCG}
     &S_{\ell k}^{(\rm tcg)}(\Delta t)=\lambda^2(\mathsf{T}^\dag(S^+(\Delta t)-S^-(\Delta t)) \mathsf{T})_{\ell k},\\
     \label{eq:discoeffCG}
     &\alpha_{\ell k}^{\pm(\rm tcg)}(\Delta t)=\frac{\lambda^2}{2}(\mathsf{T}^\dag\gamma^\pm(\Delta t)\mathsf{T})_{\ell k} \ ,
 \end{align}
where  $S^\pm(\Delta t)=[S^\pm (\Delta t)_{ij}]$ and $\gamma^\pm(\Delta t)$ are the matrices with entries the coefficients of the Lamb-shift Hamiltonian in Eq.~(\ref{lambccoeff}), respectively the coefficients of the dissipative term part in the master equation in Eq.~(\ref{eq:CG bosonic chain})
which has been defined in Eq.~(\ref{eq:gamma}).
The details of the calculation can be found in Appendix~\ref{app:Continuity equation}.
Again, as in the previous cases, also in this one the time-variation of the probability is given by the divergence of a current with unitary and dissipative contributions and without  sink/source terms. As we shall see in the next section, the appearance of such terms characterizes the continuity equation relative to the energy flow through the mid site of the chain.

\section{Energy continuity equation}
\label{sec:ECE}

While the various approaches lead to a standard continuity equation in the case of the probability flow, we now show that the picture gives a consistent 
turn when one  considers the flow of energy through the middle of the harmonic chain; namely, when focusing on the average excitation number of the middle oscillator.

We thus address the continuity equation for $\expval{a_2^\dag a_2}_t$,
\begin{eqnarray}
\nonumber
    \frac{d}{dt}\expval{a_2^\dag a_2}_t&=&\operatorname{Tr}\left(a_2^\dag a_2\frac{d}{dt}\rho(t)\right)=\operatorname{Tr}\left(a_2^\dag a_2\mathcal{L}[\rho(t)]\right)
    \\ \label{eq:mean-midd-energy}
       &=&\operatorname{Tr}\big(\mathcal{L}^{\rm adj} [a_2^\dag a_2]\rho(t)\big),
\end{eqnarray}
in the exact, global, local and time coarse-grained approaches.  

\subsection{Exact reduced dynamics}

\label{sec:continuity-equation-exact}
The exact reduced dynamics of the number of excitations of the second oscillator of the open chain is obtained by tracing out the environment from Eq.~(\ref{eq:von-neumann}) with $\hat{H}_{\rm T}$ as in Eq.~(\ref{Hamtotal}). Using Eq.~(\ref{eq:mean-midd-energy}), its mean energy evolves in time according to
\begin{align}
\label{expval4}
    \frac{d}{dt}\expval{a_2^\dag a_2}_t^{(\rm exc)} &=-g \operatorname{Tr}_{S}\left((J_{12}-J_{23})\rho^{(\rm exc)}_S(t))\right)\cr
    &=-g \expval{J_{12}-J_{23}}_t^{(\rm exc)}.
\end{align}
where $\expval{\bullet}_t^{(\rm exc)}$ denotes the expectation value of $\bullet$ with respect to $\rho^{(\rm exc)}_{S}(t)$ and $J_{jk}$ is
\begin{equation}
\label{eq:currentij}
    J_{jk}=i(a_j a_k^\dag-a_j^\dag a_k)\ .
\end{equation}
The superscript $(\rm exc)$ emphasizes that the system state is given by $\rho^{(\rm exc)}_S(t)={\rm Tr}_E(\rho_{\rm T}(t))$, where $\rho_{\rm T}(t)$ is the joint state of system and environment  evolved at time $t$ under the exact evolution generated by $\hat{H}_{\rm T}$. 

The sign of the expectation value of the current in Eq.~(\ref{eq:currentij}) provides insight into the direction of the excitation flow between sites $j$ and $k$ in the chain. Indeed, $a_ja_k^\dag$ describes the transfer of an excitation from site $j$ to site $k$, with $a_j^\dag a_k$ describing the reverse process. 
If $\expval{J_{jk}}>0$, it implies that the excitation flows from site $j$ to site $k$.
This directionality is consistent with the expected energy transport from the hotter bath to the colder bath. Being difference of currents, terms like $\expval{J_{12}-J_{23}}^{(\rm exc)}_t$ show discrete divergence-like features.

\subsection{Local approach}
\label{sec:continuity-equation-local}

In the local approach, the dissipative evolution of any observable of the open chain follows by using the so-called dual, or Heisenberg version, of the master equation in Eq.~(\ref{eq:masterlocal}); then:
\begin{align}
\label{eq:expvallocal}
    \frac{d}{dt}\expval{a_2^\dag a_2}_t^{(\rm loc)}
    &=-g \operatorname{Tr}_{S}\left((J_{12}-J_{23})\rho^{\rm (loc)}_{S}(t)\right)\cr
    &=-g \expval{J_{12}-J_{23}}_t^{(\rm loc)},
\end{align}
with $J_{jk}$ as in Eq.~(\ref{eq:currentij}).

\subsection{Global approach}
\label{sec:continuity-equation-global}

The global approach gives rise to a richer continuity equation for $a_2^\dag a_2$ than its exact and local counterparts:
\begin{align}
\label{eq:globalcountinuity}
    \frac{d}{dt}\expval{a_2^\dag a_2}_t^{(\rm glb)}&=-g'\expval{J_{12}-J_{23}}_t^{(\rm glb)}+\sum_{\alpha=L,R}\expval{Q_{\alpha}}_t^{(\rm glb)},
\end{align}
where 
\begin{equation}
    g'=\Big(g-\frac{\sqrt{2}\lambda^2}{4}(S(\epsilon_1)-S(\epsilon_3))\Big),
\end{equation}
and $S(\epsilon_i)$ and $J_{jk}$ are defined in Eq.~(\ref{eq:Sglobal}), Eq.~(\ref{eq:currentij}), respectively.
Instead, $\expval{Q_{\alpha}}_t^{(\rm glb)}$ is given by
\begin{align}
\label{eq:sinkandsource}
&\expval{Q_{\alpha}}_t^{(\rm glb)}=\frac{\sqrt{2}\pi\lambda^2}{16}\big(J_{\alpha}(\epsilon_1)-J_{\alpha}(\epsilon_3)\big)\expval{Q_{12}+Q_{23}}_t^{(\rm glb)}\cr
&-\frac{\pi\lambda^2}{4}\sum_{i=1,3}J_{\alpha}(\epsilon_i)\left(\expval{a_2^\dag a_2}_t^{(\rm glb)}-\bar{n}_{\alpha}(\epsilon_i)\right),
\end{align}
where
\begin{equation}
    \label{eq:Q-ij}
    Q_{ij}=a_i^\dag a_j+a_i a_j^\dag.
\end{equation}

It is important to notice that the right hand side of Eq.~(\ref{eq:sinkandsource}) cannot be written as the difference of two dissipative current terms $\widetilde{J}_{12}$, respectively $\widetilde{J}_{23}$ involving the oscillators $1,2$, respectively $2,3$.
Thus, $\expval{Q_{\alpha}}_t^{\rm glb}$ cannot correspond to a divergence-like term of a current and thus provides a  contribution similar to the presence of a sink if it contrives to decrease
$\expval{a^\dag_2\,a_2}_t^{(\rm glb)}$, otherwise to a source of excitations in the second oscillator.
Particularly informative about the transport properties of open many-body systems are the steady states, that are left invariant by the reduced dynamics and thus satisfy the stationarity condition $\partial_t\rho_t=\mathcal{L}[\rho_t]=0$. As sketched in Remark~\ref{rem:Gauss}, in the case of the open harmonic chain, the various dissipative generators exhibit each a unique Gaussian steady state completely specified by its own asymptotic covariance matrix.
Within the framework of the global approach to this model, the steady state of the harmonic chain is characterized by the covariance matrix $C_{\rm glb}^\infty$, which is~\cite{babakan_open_2025}:
\begin{align}
\label{eq:C1globalGeneral}
   &C^\infty_{\rm glb}=\mathsf{T}\widetilde{C}_1\mathsf{T}^\top\oplus \mathsf{T}\widetilde{C}_1^\top\mathsf{T}^\top,\\
   &\widetilde{C}_1=\sum_{\alpha=L,R}\operatorname{diag}\big(\bar{n}_\alpha(\epsilon_1),\bar{n}_\alpha(\epsilon_2),\bar{n}_\alpha(\epsilon_3)\big)+\mathbb{I}_3, 
\end{align}
where $\mathbb{I}_3$ denotes a $3$-dimensional identity matrix and
\begin{equation}
   \label{eq:CovGen}C_{ij}=\langle\xi_i\xi_j^\dag+\xi_j^\dag\xi_i\rangle-2\langle\xi_i\rangle\langle \xi_j^\dag\rangle,
\end{equation} 
with $\hat{\xi}=(a_1,a_2,a_3,a_1^\dag,a_2^\dag,a_3^\dag)^\top$. Using the covariance matrix of the steady state in the global approach in Eq.~(\ref{eq:C1globalGeneral}), one computes that both currents $J_{12}$ and $J_{23}$ have vanishing mean values with respect to the steady state:
$ \expval{J_{12}}_\infty^{(\rm glb)}=\expval{J_{12}}_\infty^{(\rm glb)}=0$.
Additionally, one finds that sink and source terms read
\begin{align}
\label{eq:Ql}
    \expval{Q_L}_\infty^{(\rm glb)}&=\frac{\pi}{2}\lambda^2\sum_{i=1,3}(\bar{n}_{L}(\epsilon_i)-\bar{n}_R(\epsilon_i)),\\
    \label{eq:Qr}
    \expval{Q_{R}}_\infty^{(\rm glb)}&=\frac{\pi}{2}\lambda^2\sum_{i=1,3}(\bar{n}_R(\epsilon_i)-\bar{n}_L(\epsilon_i))\ .
\end{align}
Thus, the stationary mean-values of the left and right sink and source terms compensate each-other; namely,
$\expval{Q_L}_\infty^{(\rm glb)}=-\expval{Q_R}_\infty^{(\rm glb)}$.

Comparing the exact and local approaches, the continuity equation of $\expval{a_2^\dag a_2}_t$ shows an identical structure   (Eqs.~(\ref{expval4}) and (\ref{eq:expvallocal})), which contains a standard current divergence-like term.
In contrast, the global approach introduces an additional contribution in Eq.~(\ref{eq:globalcountinuity}) that cannot be expressed as the difference of two currents and cannot then but be identified as a composite source/sink term.
Notably, while the exact derivation requires no approximations, both the local and global approaches employ simplifying assumptions. In the following, we will
study which specific approximations lead to sink/source terms in the global continuity equation for the expectation value of the central oscillator's number operator.
Still working in the interaction picture, applying the adjoint of the WCL generator, $\mathcal{L}_t[\bullet]$, in Eq.~(\ref{eq:weak-coupling-Gen}) to Eq.~(\ref{eq:mean-midd-energy}) yields:
\begin{equation}
\frac{d}{dt}\widetilde{\expval{a_2^\dag a_2}}_t^{(\rm wcl)}= \operatorname{Tr}\left(\mathcal{L}_{t}^{\rm adj}[a_2^\dag a_2]\tilde{\rho}(t)\right) = 0\ ,
\end{equation}
where $\widetilde{\expval{X}_t}$ denotes the mean value of the observable $X$ taken relative to the state $\widetilde{\rho}_t$ evolved up to time $t$ in the interaction representation.
Therefore, moving back to the Schrödinger picture gets a divergence-like contribution to the continuity equation, with no source/sink terms emerging at this level of approximation.

Similarly, employing the adjoint of the time-rescaling generator, $\mathcal{L}_{\rm tr}[\bullet]$ in Eq.~(\ref{eq:time-rescaling-Gen}), one obtains
\begin{equation}
\frac{d}{dt}\widetilde{\expval{a_2^\dag a_2}}_t^{(\rm tr)} = \operatorname{Tr}\left(\mathcal{L}_{\rm tr}^{\rm adj}[a_2^\dag a_2]\tilde{\rho}(t)\right) = 0,
\end{equation}
thereby showing that this further approximation also does not introduce source/sink terms.

The final approximation to achieve a Markovian master equation is the rotating wave-approximation which amounts to using the generator $\mathcal{L}[\bullet]$ in Eq.~(\ref{eq:secular-approximation-Gen}). Then,
\begin{align}
\frac{d}{dt}\widetilde{\expval{a_2^\dag a_2}}_t^{(\rm RWA)} &= \operatorname{Tr}\left(\mathcal{L}^{\rm adj}[a_2^\dag a_2]\tilde{\rho}(t)\right)\cr
&=\sum_{\alpha=\ell,r}\widetilde{\expval{Q_{\alpha}}}_t^{(\rm glb)} \neq 0,
\end{align}
where the right-hand quantity is defined as in Eq.~(\ref{eq:sinkandsource})) and constitutes a source/sink contribution. This shows that it is at the level of the RWA that non-divergence-like terms emerge in the global approach.
As pointed out in Appendix~\ref{CG}, the time-coarse-graining approach on one hand does not use the RWA and, on the other hand, 
permits the inspection of different time-scales through the time-zoom parameter $\Delta t$; we then discuss the energy continuity equation in the case of the coarse-grained master equation derived in \S~\ref{sec:coarse-grained-master-equation}.

\subsection{Time-coarse-graining approach}

\label{sec:continuity-equation-CG}

In the time-coarse-graining approach, from~\eqref{eq:CG bosonic chain}, one gets the following continuity equation for 
$a_2^\dag a_2$:
\begin{equation}
\label{eq:contiEqTcg}
\frac{d}{dt}\expval{a_2^\dagger a_2}^{(\rm tcg)}_t \hskip-.5cm =   A(\Delta t) \expval{J_{12} - J_{23}}^{(\rm tcg)}_t
\hskip-.2cm + \hskip-.1cm \sum_{\alpha=L,R} \expval{Q_\alpha}^{(\rm tcg)}_t,
\end{equation}
where the coefficient $A(\Delta t)$ is
\begin{eqnarray}
\nonumber
    A(\Delta t) &=&(\mathsf{T}^\dag\Xi\mathsf{T})_{12}+\Im(\bar{\alpha}^{+(\rm tcg)}(\Delta t)-\alpha^{-(\rm tcg)}(\Delta t))_{12}\\
    \label{eq:coffA}
    &-&\Re(S_{12}^{(\rm tcg)}(\Delta t)),
\end{eqnarray}
with $\Xi=\operatorname{diag}(\epsilon_1,\epsilon_2,\epsilon_3)$, $S_{\ell k}^{(\rm tcg)}(\Delta t)$ and $\alpha^{\pm(\rm tcg)}(\Delta t)_{\ell k}$ in Eqs.~(\ref{eq:SlambCG}) and (\ref{eq:discoeffCG}), respectively.
As in the probability continuity equation for the  time-coarse-grained reduced dynamics (see Eq.~\eqref{eq:dissconteq2}), where the master equation also provides  dissipative currents, similarly do the second and third terms in Eq.~(\ref{eq:coffA}). In contrast, the last term on the r.h.s of Eq.~(\ref{eq:contiEqTcg}) is  a sink/source term which can not be recast as the difference of two currents; indeed,
\begin{eqnarray*}
    \expval{Q_\alpha}^{(\rm tcg)}&=&B(\Delta t)\expval{Q_{12}+Q_{23}}^{(\rm tcg)}_t+D(\Delta t)\expval{a_2^\dag a_2}^{(\rm tcg)}_t\\
    &+&F(\Delta t),
\end{eqnarray*}
with coefficients 
\begin{eqnarray*}
    B(\Delta t)&=&-\frac{\sqrt{2}\pi\lambda^2}{32}\Delta t\int_{-\omega_c}^{\omega_c} d\omega \, J_\alpha(\omega)\,\times\\
    &\times&\left( \operatorname{sinc}^2\left((\omega - \epsilon_3) \frac{\Delta t}{2}\right) - \operatorname{sinc}^2\left((\omega - \epsilon_1) \frac{\Delta t}{2}\right) \right)\ ,
    \end{eqnarray*}
\begin{eqnarray*}
    D(\Delta t)&=&-\frac{\lambda^2\pi}{8}\Delta t \int_{-\omega_c}^{\omega_c} d\omega \, J_\alpha(\omega)\,\times\\
    &\times&\left( \operatorname{sinc}^2\left((\omega - \epsilon_3) \frac{\Delta t}{2}\right) + \operatorname{sinc}^2\left((\omega - \epsilon_1) \frac{\Delta t}{2}\right) \right)\\
    &-&\frac{\sqrt{2}\pi\lambda^2}{2}\Delta t \cos{(\sqrt{2}g \Delta t)} \int_{-\omega_c}^{\omega_c} d\omega J_\alpha(\omega)\bar{n}_\alpha(\omega)\\
    &\times&\operatorname{sinc}\left((\omega - \epsilon_1) \frac{\Delta t}{2}\right) \operatorname{sinc}\left((\omega - \epsilon_3) \frac{\Delta t}{2}\right)\ ,
    \end{eqnarray*}
    \begin{eqnarray*}
    F(\Delta t)&=& \frac{\lambda^2\pi}{8}\Delta t \int_{-\omega_c}^{\omega_c} d\omega \, J_\alpha(\omega)\bar{n}_\alpha(\omega)\,\times\\
    &\times&\left( \operatorname{sinc}^2\left((\omega - \epsilon_3) \frac{\Delta t}{2}\right) + \operatorname{sinc}^2\left((\omega - \epsilon_1) \frac{\Delta t}{2}\right) \right)\\
    &-&\frac{\sqrt{2}}{2}\pi\lambda^2\Delta t \cos{(\sqrt{2}g \Delta t)} \int_{-\omega_c}^{\omega_c} d\omega J_\alpha(\omega)n_\alpha(\omega)\,\times\\
    &\times&\operatorname{sinc}\left((\omega - \epsilon_1) \frac{\Delta t}{2}\right) \operatorname{sinc}\left((\omega - \epsilon_3) \frac{\Delta t}{2}\right)\ ,
\end{eqnarray*}
where $J_{nm}$ and $Q_{mn}$ are defined in Eq.~(\ref{eq:currentij}) and Eq.~(\ref{eq:Q-ij}).

It proves now convenient to briefly examine the energy  transport properties in the asymptotic regime, namely with respect to the unique stationary Gaussian state $\rho^{\rm tcg}_\infty$ of the reduced dynamics (see Remark~\ref{rem:Gauss}) such that $\partial_t\expval{a_2^\dagger a_2}^{(\rm tcg)}_{t}=0$. As we know, $\rho^{(\rm tcg)}_\infty$ is identified by its covariance matrix. Then, the relevant currents, $\expval{J_{12}}^{(\rm tcg)}_{\infty}$ and $\expval{J_{23}}^{(\rm tcg)}_{\infty}$, and source/sink terms, $\expval{Q_L}^{(\rm tcg)}_{\infty}$ and $\expval{Q_{R}}^{(\rm tcg)}_{\infty}$, are functionals of the covariance matrix $C$, defined in Eq.~(\ref{eq:CovGen}) (See Appendix~\ref{app:CovTcg} for more detail).
\begin{figure}
    \centering
    \includegraphics[width=1\linewidth]{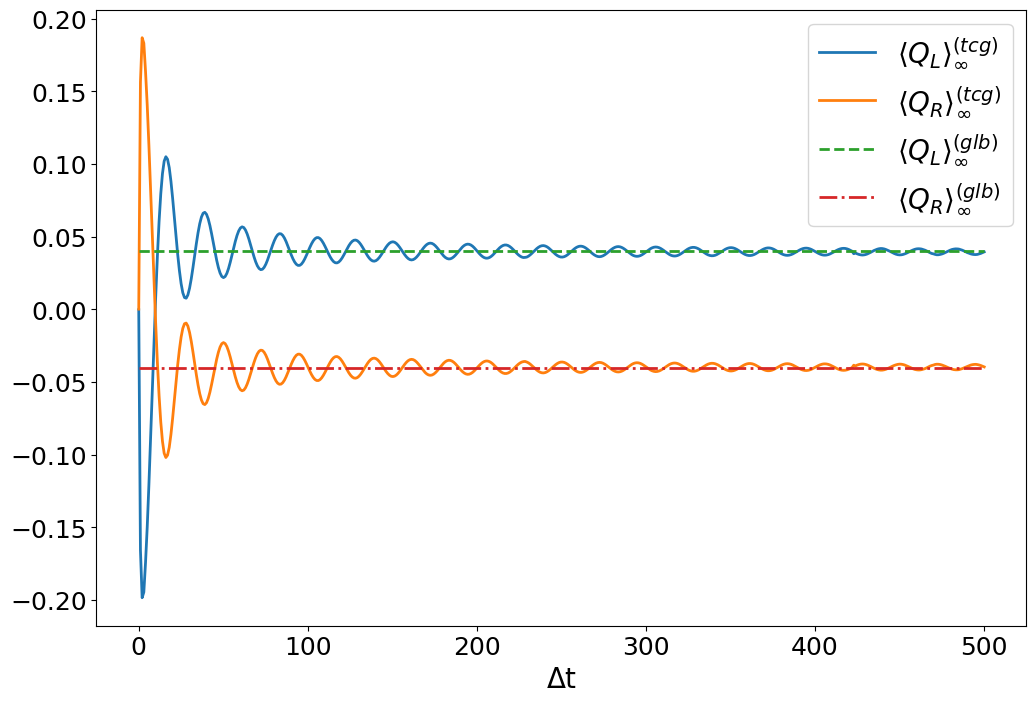}
    \caption{Here, we plot $\expval{Q_\alpha}_\infty^{(\beta)}$ for $\alpha=L,R$ and $\beta=\rm{tcg},\rm{glb}$ with the steady state $\rho_\infty^{(\rm tcg)}$ and $\rho_\infty^{(\rm glb)}$, respectively. The blue and orange lines are $\expval{Q_L}_\infty^{(\rm{tcg})}$ and $\expval{Q_R}_\infty^{(\rm{tcg})}$, while the green dashed line and red dotted-dashed lines are $\expval{Q_L}_\infty^{(\rm{glb})}$ and $\expval{Q_R}_\infty^{(\rm{glb})}$, respectively. We set $\lambda=0.1$, $\omega_0=1$, $\omega_c=3$, $T_L=10$ and $T_R=1$.}
    \label{fig:QLQRGlbTcg}
\end{figure}
As can be seen in Fig.~\ref{fig:QLQRGlbTcg}, $\expval{Q_L}_{\infty}$ and $\expval{Q_R}_{\infty}$ have been plotted for both the global and the time-coarse-graining approaches, alongside their respective steady states. As the coarse-graining time scale $\Delta t \to \infty$, the values of $\expval{Q_L}_{\infty}^{(tcg)}$ and $\expval{Q_R}_{\infty}^{(tcg)}$ approach their global counterparts $\expval{Q_L}_{\infty}^{(glb)}$ and $\expval{Q_R}_{\infty}^{(glb)}$ . This convergence provides further evidence that the time-coarse-graining formalism reproduces the results of the global GKSL master equation in the long-time limit ($\Delta t \to \infty$). Conversely, in the limit of $\Delta t \to 0$, both $\expval{Q_L}_{\infty}^{(tcg)}$ and $\expval{Q_R}_{\infty}^{(tcg)}$ vanish, indicating the absence of any source or sink terms on the finest, that is instantaneous  time scale. In the steady state of either the time-coarse-graining approach or the 
global one, the currents $\expval{J_{12}}_{\infty}$ and $\expval{J_{23}}_{\infty}$ are also zero, and $\expval{Q_L}= -\expval{Q_R}$ satisfies the stationarity condition.

\section{conclusions}
In this work, we have systematically analyzed particle and energy transport in an open quantum system comprising a three harmonic oscillator chain coupled to thermal baths at different temperatures at the ends of the chain. By comparing exact, local, and global master equation approaches, we demonstrated that while all three frameworks yield a divergence-like continuity equation for the probability current, they exhibit marked differences in their description of energy transport. Specifically, the global approach, relying on the rotating wave approximation (RWA), introduces non-trivial sink/source terms in the energy continuity equation that cannot be expressed as a discrete divergence-like difference of currents. In contrast, the exact and local approaches preserve a standard divergence-like form.

By using the time-coarse-graining method that circumvents the RWA, we showed that these anomalous terms though vanishing in the limit of fine time resolution ($\Delta t\to 0$), are still present. This behavior, on one hand reconciles the global approach’s predictions with the exact dynamics; however, it also indicates that sink and source terms are not ascribable solely to the RWA. This result highlights the critical role of approximation choices in modeling energy transport and underscores how collective effects, even in weakly interacting systems, can significantly influence transport properties when interactions are neglected in the coupling to the environment.

Our findings contribute to the ongoing debate about the reliability of local versus global open reduced  dynamics, emphasizing the need of a particular care  and consideration of the physical time-scales when approximating the time-evolution of open quantum systems. Indeed, on the time-scales where a global reduced dynamics is preferable, sink and source terms would affect the energy flow. This effects are instead not be visible on the time-scale proper to a local reduced dynamics.

Future work could extend this analysis to larger systems scenarios, where the interplay between interactions and dissipation may lead to yet a richer phenomenology. Ultimately, the previous considerations and results indicate how open harmonic chains might provide quite a useful setting for assessing 
the validity of common approximations in quantum transport problems.

\begin{acknowledgments}
  F.B. acknowledges financial support from PNRR MUR project PE0000023-NQSTI. L.M. and M.B acknowledge financial support from the Iran National Science Foundation (INSF) under Project No. 4022322. L. M. acknowledges support from the ICTP through the Associates Programme (2019-2024). M. B. acknowledges financial support from International Center for Theoretical Physics (ICTP) under the Sandwich Training Education Program (STEP). 
\end{acknowledgments}

\appendix

\section{GKLS master equation}
\label{app1}

The standard weak-coupling limit setting consists of a system $S$ in weak interaction with an environment $E$  via a total Hamiltonian as in Eq.~(\ref{Hamtotal}) with interaction of the form 
\begin{equation}
\label{eq:intHamiltonian}
    \hat{H}'=\sum_iA_i\otimes B_i ,
\end{equation}
where $A_i$ and $B_i$ are system and bath operators, respectively.
Furthermore, one considers an initial factorized compound state $\rho_S(0)\otimes\rho_E$, where $\rho_E$ is a thermal equilibrium environment state.
In order to derive a master equation of the GKLS form by eliminating the environment degrees of freedom, it proves convenient  to work in the interaction
representation:
\begin{equation}
\tilde\rho_{_{\rm T}}(t)=e^{it\hat{H}_0}\ \rho_{_{\rm T}}(t)\ e^{-it\hat{H}_0}\ ,\quad \hat{H}_0=\hat{H}_S+\hat{H}_E.
\label{6}
\end{equation}
Setting $\displaystyle \hat{H}'(t)=e^{it\hat{H}_0}\ \hat{H}'\ e^{-it\hat{H}_0}$,
\begin{equation}
\frac{d}{dt}\tilde\rho_{_{\rm T}}(t)=-i \lambda\Big[\hat{H}'(t),\, \tilde\rho_{_{\rm T}}(t)\Big].
\label{7}
\end{equation}
The so-called W(eak) C(oupling) L(imit) procedure consists in three steps: in the first one expresses the solution to~\eqref{7} as a Dyson expansion at second order in the small parameter $\lambda$ so that  the system density matrix  in interaction picture, $\Tilde{\rho}(t)={\rm Tr}_E\Tilde{\rho}_{_{\rm T}}(t)$, is approximated by
\begin{eqnarray}
\nonumber
&&
\Tilde{\rho}(t)=\rho(0)\\
\nonumber       
&&
+\lambda^2\int_{0}^tds\sum_{\omega\omega'}e^{i(\omega'-\omega)s}\Gamma_{kl}^s(\omega)[A_l(\omega)\Tilde{\rho}(s),A_k^\dag(\omega')]\\
&&
+\lambda^2\int_{0}^tds\sum_{\omega\omega'}e^{-i(\omega'-\omega)s}\Gamma_{lk}^{s*}(\omega)[A_l(\omega'),\Tilde{\rho}(s)A_k^\dag(\omega')],
       \label{eq:rho-weak-coupling}\\
\nonumber
&&\hbox{where}\qquad \Gamma_{kl}^s(\omega)=\int_0^sdue^{i\omega u}\operatorname{Tr}_E\Big(\Tilde{B}_k(s)\Tilde{B}_l(s-u)\rho_E\Big).
\end{eqnarray}
From Eq.~(\ref{eq:rho-weak-coupling}) we thus get a time-dependent generator
\begin{equation}
       \frac{d}{dt}\tilde{\rho}_S(t):=\mathcal{L}_t[\tilde{\rho}_S(t)] 
\end{equation}
with
\begin{align}
\label{eq:weak-coupling-Gen}
       \mathcal{L}_t[\bullet]&=\lambda^2\sum_{\omega\omega'}e^{i(\omega'-\omega)t}\Gamma_{kl}^t(\omega)[A_l(\omega)\bullet,A_k^\dag(\omega')]\cr
       &+\lambda^2\sum_{\omega\omega'}e^{-i(\omega'-\omega)t}\Gamma_{lk}^{t*}(\omega)[A_l(\omega),\bullet A_k^\dag(\omega')]\ .
\end{align}
In order to get the time-independent generator of a Markovian semigroup dynamics, the second step of the WCL approximation
consists in looking at the open system dynamics given by~\eqref{eq:rho-weak-coupling} on a long physical time-scale obtained 
by going to a rescaled time $\tau=\lambda^2\,t$ and letting $\lambda\to0$. The system density matrix in the interaction picture after re-scaling the time is
\begin{align}
\label{eq:time-rescaling-Gen}
       &\Tilde{\rho}(t)=\rho(0)\cr
       &+\lim_{\lambda\to 0}\sum_{\omega\omega'}\int_{0}^\tau d\sigma\Omega_{kl}(\omega,\omega',\frac{\sigma}{\lambda^2})[A_l(\omega)\Tilde{\rho}(\sigma/\lambda^2),A_k^\dag(\omega')]\cr
       &+\lim_{\lambda\to 0}\sum_{\omega\omega'}\int_{0}^\tau d\sigma\Omega_{kl}(\omega,\omega',\frac{\sigma}{\lambda^2})^*[A_l(\omega'),\Tilde{\rho}(\sigma/\lambda^2)A_k^\dag(\omega')],\cr
\end{align}
with 
\begin{equation}
    \Omega_{kl}(\omega,\omega',\frac{\sigma}{\lambda^2}) = e^{i(\omega'-\omega)\sigma/\lambda^2}\Gamma_{kl}^{\sigma/\lambda^2}(\omega)\ .
\end{equation}
The final step in the WCL approximation is implementing the secular approximation (SA) which consists in eliminating all fast oscillations by keeping only terms with $\omega'=\omega$. Then,
   \begin{align}
       &\Tilde{\rho}(\tau)=\rho(0)\cr
       &+\lambda^2\int_{0}^\infty d\sigma\sum_{\omega}\Gamma_{kl}^{\infty}(\omega)[A_l(\omega)\Tilde{\rho}(\sigma),A_k^\dag(\omega)]\cr
       &+\lambda^2\int_{0}^\infty d\sigma\sum_{\omega}\Gamma_{lk}^{\infty*}(\omega)[A_l(\omega),\Tilde{\rho}(\sigma)A_k^\dag(\omega)].
   \end{align}
At the end of the procedure, the WCL yields the following Markovian master equation with time-independent GKSL generator:
\begin{align}
\label{eq:secular-approximation-Gen}
    \dot{\rho}(t)&=\mathcal{L}[\rho(t)]=-i[\hat{H}_S+\lambda^2 \hat{H}_{\rm LS},\rho(t)]\cr
    &+\sum_{i,k}(A_i(\omega_k)\rho(t)A_i^\dag(\omega_k)-\frac{1}{2}\{A_i^\dag(\omega_k)A_i(\omega_k),\rho(t)\}),\cr
\end{align}
where $\hat{H}_{\rm LS}$ is the Lamb-shift Hamiltonian
\begin{eqnarray}
\label{HLS}
    \hat{H}_{\rm LS}&=&\sum_{i,j,k}S_{ij}(\omega_k)A_j^\dag(\omega_k)A_i(\omega_k)\quad\hbox{and}\\
\label{SOp}
    S_{ij}(\omega_k)&=&\mathcal{P}.\mathcal{V}.\int_{-\omega_{max}}^{\omega_{\max}} d\nu \frac{\operatorname{Tr}\left(B_j^\dag(\nu)B_i\rho_B\right)}{\omega_k-\nu},
\end{eqnarray}
with $\mathcal{P}.\mathcal{V}.$ denoting the principal value of the integral.
The Lindblad operators $A_i(\omega_k)$ are instead given by
\begin{equation}
\label{eq:LindOp}
    A_i(\omega_k)=\sum_{\epsilon-\epsilon'=\omega_k}P_{\epsilon'}\,A_i\,P_{\epsilon}\ ,
\end{equation}
where $P_{\epsilon}$ are the projection operators onto the eigenstates of $\hat{H}_S$ while the operators $A_i$ appear in the interaction Hamiltonian~(\ref{eq:intHamiltonian}). 

\section{Derivation of  the time-coarse-graining master equation}
\label{CG}

Consider the environment induced Lamb-shift Hamiltonian and dissipative contribution in Eqs.~(\ref{eq:HLambTCG}--\ref{13}), respectively:
\begin{align*}
  \hat{H}^{({\rm LS})}_{\rm tcg}(\Delta t)&={i\lambda^2\over2\Delta t}\int_0^{\Delta t}ds_1\int_0^{\Delta t}
\theta(s_1-s_2)\cr
&\times {\rm Tr}_E\Big(\rho_E\big[ \tilde{H}'(s_1),\tilde{H}'(s_2)\big]\Big),\\
 {\cal D}_{\rm tcg}[\rho_S(t)]&={\lambda^2\over \Delta t} {\rm Tr}_E\Big(
L(\Delta t)\,\big(\rho_S(t)\otimes\rho_E\big)\, L(\Delta t)\cr
&-{1\over2}\Big\{ L^2(\Delta t),\big(\rho_S(t)\otimes\rho_E\big)\Big\}\Big),
\end{align*}
where $L(\Delta t)=\int_0^{\Delta t}ds \tilde{H}'(s)$. 
We first concentrate on the dissipative contribution 
\begin{eqnarray}
\nonumber
    {\cal D}_{\rm tcg}[\rho_S(t)]&=&{\lambda^2\over \Delta t}\int_0^{\Delta t}\hskip-.5 cmds\int_0^{\Delta t}\hskip-.5cm ds'{\rm Tr}_E\Bigg(\tilde{H}'(s)\,\big(\rho_S(t)\otimes\rho_E\big)\, \tilde{H}'(s')\\
\label{eq:A3}   &-&{1\over2}\Big\{ \tilde{H}'(s')\tilde{H}'(s),\big(\rho_S(t)\otimes\rho_E\big)\Big\}\Bigg)\ .
\end{eqnarray}
Before dealing with the Hamiltonians in Eqs.~(\ref{HS}--\ref{Hint}), in the interaction picture we set
$$
\tilde{H}'(s)=e^{is\hat{H}_0}\ \hat{H}'\ e^{-is\hat{H}_0}\ ,\\
$$
where $\hat{H}_0=\hat{H}_S+\hat{H}_E$, $\hat{H}'=\sum_k A_k\otimes B_k$ and 
$A_k$ and $B_k$ hermitian operators acting on the Hilbert space of the system and environment, respectively.
Given the eigenvectors and eigenvalues of $\hat{H}_S$, $\hat{H}_S\vert \varepsilon_j\rangle=\varepsilon_j \vert \varepsilon_j\rangle$, we let $\omega$ run over the possible transition frequencies $\varepsilon_i-\varepsilon_j$ and set 
\begin{equation}
\label{App1:lindb-ops}
\hskip-.1cm
A_k(\omega)=\hskip-.3cm\sum_{\varepsilon_i-\varepsilon_j=\omega}\hskip-.3cm\vert\varepsilon_j\rangle\langle\varepsilon_j\vert A_k\vert\varepsilon_i\rangle\langle\varepsilon_i\vert.%\ B_k=\hskip-.1cm\int_{-\infty}^{\infty}\hskip-.3cmd\omega B_k(\omega) \ .
\end{equation}
Therefore, in the interaction picture the interaction Hamiltonian explicitly reads
\begin{eqnarray}
\nonumber
\tilde{H}'(s)&=&\sum_k e^{is\hat{H}_S}\ A_k\ e^{-is\hat{H}_S} \otimes \underbrace{e^{is\hat{H}_E}\ B_k\ e^{-is\hat{H}_E}}_{B_k(s)}\\
\label{lind-ops}
&=&\sum_k\Big(\sum_\omega e^{is\omega} A_k(\omega) \otimes B_k(s)\Big),
\end{eqnarray}
while the generator in Eq.~(\ref{eq:A3}) becomes
\begin{eqnarray}
\nonumber
&&\hskip -.9cm
 {\cal D}_{\rm tcg}[\rho_S(t)]={\lambda^2\over \Delta t}\sum_{\omega_1,\omega_2}\sum_{k,\ell}\int_0^{\Delta t}\hskip-.3cm
    ds\int_0^{\Delta t}\hskip -.3cm ds'\times\\
    \nonumber
&&\hskip -.9cm
\times e^{i\omega_1s}e^{-i\omega_2s'}\,{\rm Tr}_E\Big(\rho_E B_k(s)B_\ell(s^\prime)\Big)\times\\
\label{eq:A6}    
&&\hskip -0.9cm
\times\Bigg(\hskip-.1cm A_k(\omega_1)\rho_S(t) A_\ell^\dag(\omega_2)-{1\over2}\Big\{A_\ell^\dag(\omega_2)A_k(\omega_1) ,\rho_S(t)\Big\}\hskip-.1cm\Bigg).
\end{eqnarray}
It proves convenient to introduce the Fourier transform of the two-point time-correlation function:
\begin{equation}
\label{App1:correlations1}
{\rm Tr}_E\Big(\rho_E B_k(s)B_\ell(s^\prime)\Big)={\rm Tr}_E\Big(\rho_E B_k(s-s^\prime)B_\ell\Big),
\end{equation}
where the time-translation invariance of the environment state $\rho_E$ has been used. Namely,
\begin{equation}
\label{App1:correlations2}
\sigma_{k\ell}(\omega):=\frac{1}{2\pi}\int_{-\infty}^{+\infty}ds\,e^{-i\omega s}{\rm Tr}_E\Big(\rho_EB_k(s)B_\ell\Big),
\end{equation}
so that
\begin{equation}
\label{App1:correlations3}
{\rm Tr}_E\Big(\rho_EB_k(s)B_\ell\Big)=\int_{-\infty}^{+\infty}d\omega\,e^{i\omega s}\,\sigma_{k\ell}(\omega).
\end{equation}
\begin{remark}
\label{rem3}
Typically, the environment spectral density is characterized by a cut-off frequency $\omega_c$ which allows us to approximate
\begin{equation}
\label{correlations3}
{\rm Tr}_E\Big(\rho_EB_k(s)B_\ell\Big)\simeq\int_{-\omega_c}^{+\omega_c}d\omega\,e^{i\omega s}\,\sigma_{k\ell}(\omega)\ .
\end{equation}
The coarse-graining time scale $\Delta t$ corresponds to the time-scale after which the environment has effectively reset to its equilibrium state after interacting with the harmonic chain. It must then satisfy the bounds $\tau_c\ll \Delta t \ll \tau_s$,
where $\tau_c$ is the inverse of $\omega_c$ and $\tau_s$ is the time-scale for significant changes in the system state to occur.
\end{remark}

Inserting Eq.~(\ref{App1:correlations3}), into Eq.~(\ref{eq:A6}) one gets
\begin{eqnarray}
\nonumber
&&\hskip -.9cm
 {\cal D}_{\rm tcg}[\rho_S(t)]={\lambda^2\over 2\pi\Delta t}\sum_{\omega_1,\omega_2}\sum_{k,\ell}\int_0^{\Delta t}\hskip-.3cm
    ds\int_0^{\Delta t}\hskip -.3cm ds'\int_{-\infty}^{\infty}d\omega\times\\
    \nonumber
&&\hskip -.9cm
\times e^{i(\omega_1+\omega)s}e^{-i(\omega_2+\omega)s'}\,\sigma_{kl}(\omega)\times\\
\label{eq:A6a}    
&&\hskip -.9cm
\times\Bigg(\hskip -.1cm A_k(\omega_1)\rho_S(t) A_\ell^\dag(\omega_2)-{1\over2}\Big\{A_\ell^\dag(\omega_2)A_k(\omega_1) ,\rho_S(t)\Big\}\hskip -.2cm \Bigg).
\end{eqnarray}
while the terms of the form in Eq.~(\ref{eq:A6}) can be written as
$$
\int_0^{\Delta t} ds\,e^{i\alpha s}= \Delta t\,e^{i\alpha \frac{\Delta t}{2}}\,\operatorname{sinc}(\alpha\frac{\Delta t}{2}),
$$
where $\displaystyle\operatorname{sinc}(x)=\frac{\sin{(x)}}{x}$.
Finally, Eq.~(\ref{eq:A6}) reads
\begin{eqnarray}
\nonumber
\hskip-1.5cm
&&    {\cal D}_{\rm tcg}[\rho_S(t)]=\sum_{k\ell}\sum_{\omega_1,\omega_2}\gamma_{k,\ell}^{(\Delta t)}(\omega_1,\omega_2)\ \times\\
\label{eq:A8a}    
&&
\hskip-1.1cm
\times
\Bigg(A_k(\omega_1)\rho(t) A_\ell^\dag(\omega_2)-{1\over2}\Big\{A_\ell^\dag(\omega_2)A_k(\omega_1) ,\rho(t)\Big\}\Bigg).
\end{eqnarray}
The coefficient, $\gamma_{k,\ell}^{(\Delta t)}(\omega_1,\omega_2)$ in Eq.~(\ref{eq:A8a}) is given by
\begin{eqnarray}
\nonumber
&&    \gamma_{k,\ell}^{(\Delta t)}(\omega_1,\omega_2)=\frac{\lambda^2 \Delta t}{2\pi}\int_{-\infty}^{+\infty}d\omega
    e^{i(\omega_1-\omega_2) \frac{\Delta t}{2}}\sigma_{k\ell}(\omega)\times\\
\label{eq:A8b}    
&&\hskip .5cm
    \times\operatorname{sinc}\big ((\omega+\omega_1)\frac{\Delta t}{2}\big )\operatorname{sinc}\big ((\omega+\omega_2)\frac{\Delta t}{2}\big ).
\end{eqnarray}

In the same vein, the Lamb-shift term can be  recast as
\begin{eqnarray}
\label{Lamb-shift1}
\hskip-1cm&&
\hat{H}^{({\rm LS})}_{\rm tcg}(\Delta t)=\sum_{k\ell}\sum_{\omega_1,\omega_2}S^{\Delta t}_{k\ell}(\omega_1,\omega_2)\,A^\dag_{k}(\omega_1)\,A_{\ell}(\omega_2)\ ,\\
\nonumber
\hskip-.8cm
&&
S^{(\Delta t)}_{k\ell}(\omega_1,\omega_2)={i\lambda^2\over2\Delta t}\int_0^{\Delta t}\hskip-.3cm ds\int_0^{\Delta t} \hskip -.3cm ds^{\prime}\, 
\theta(s-s^{\prime})\times\\
 \label{Lamb-shift2}
\hskip-.8cm
&&
\times e^{is\omega_1-is^{\prime}\omega_2}\sigma_{k,\ell}(\omega)\ .
\end{eqnarray}
where 
\begin{eqnarray}
\nonumber
&&
S^{(\Delta t)}_{k\ell}(\omega_1,\omega_2)={\lambda^2\Delta t\over4\pi}e^{i(\omega_1-\omega_2)\frac{\Delta t}{2}}\sum_{k,\ell}\int_{-\infty}^{+\infty}\hskip-.3cm d\omega\sigma_{k\ell}(\omega)\times\\
\label{eq:A8c}
&&\hskip 0.1cm
\times\left (\operatorname{sinc}\big ((\omega_1-\omega_2)\frac{\Delta t}{2}\big )-\operatorname{sinc}\big ((\omega-\omega_2)\frac{\Delta t}{2}\big )\right ).
\end{eqnarray}

In the case of the harmonic chain with Hamiltonian as defined in Eq.~(\ref{HS}--\ref{Hint}), the operators 
in the environment-system  interaction term $\hat{H}'=\sum_kA_k\otimes B_k$ are
\begin{align}
    A_1=a_1,\quad B_1=\sum_k\gamma_{k\rm L}b^\dag_{k\rm L},\cr
    A_2=a_3,\quad B_2=\sum_k\gamma_{k\rm R}b^\dag_{k\rm R},\cr
    A_3=a_1^\dag,\quad B_3=\sum_k\gamma_{k\rm L}b_{k\rm L},\cr
    A_4=a_3^\dag,\quad B_4=\sum_k\gamma_{k\rm R}b_{k\rm R},
\label{opB}
\end{align}
therefore, using Eq.~(\ref{App1:lindb-ops}), the Lindblad operators read
\begin{align}
\label{A1}
    A_1(\omega)&=\sum_{E_{\vec{n}}-E_{\vec{m}}=\omega}\vert \vec{m}\rangle\langle\vec{m}\vert a_1\vert\vec{n}\rangle\langle\vec{n}\vert,\\
    A_2(\omega)&=\sum_{E_{\vec{n}}-E_{\vec{m}}=\omega}\vert \vec{m}\rangle\langle\vec{m}\vert a_3\vert\vec{n}\rangle\langle\vec{n}\vert,\\
    A_3(\omega)&=\sum_{E_{\vec{n}}-E_{\vec{m}}=\omega}\vert \vec{m}\rangle\langle\vec{m}\vert a_1^\dag\vert\vec{n}\rangle\langle\vec{n}\vert,\\
    \label{A4}
    A_4(\omega)&=\sum_{E_{\vec{n}}-E_{\vec{m}}=\omega}\vert \vec{m}\rangle\langle\vec{m}\vert a_3^\dag\vert\vec{n}\rangle\langle\vec{n}\vert,,
\end{align}
where the vectors $\vert \vec{n}\rangle$ are the eigen-vectors of $\hat{H}_S$ with eigenvalues $E_{\vec{n}}=\sum_{j=1}^3n_j\varepsilon_j$ such that 
\begin{equation}
\hat{H}_S\ket{\vec{n}}=\sum_{j=1}^3n_j\varepsilon_j\,\ket{\vec{n}}, 
\end{equation}
where $n_i$s are natural numbers. Further, writing $a_1$ and $a_3$ as functions of the operators $c_i$ in Eq.~(\ref{eq:xiglb}), 
\begin{align}
    a_1&=\frac{1}{2}(c_1+\sqrt{2}c_2+c_3),\\
    a_3&=\frac{1}{2}(c_1-\sqrt{2}c_2+c_3),
\end{align}
the Lindblad operators in Eq.~(\ref{A1}--\ref{A4}) read
\begin{align}
\label{eq:A1cc}
   A_1(\omega)&=A_4^\dag(\omega)=\frac{1}{2}(\delta_{\omega,\varepsilon_1}c_1+\sqrt{2}\delta_{\omega,\epsilon_2}c_2+\delta_{\omega,\varepsilon_3}c_3),\\
   A_2(\omega)&=A_5^\dag(\omega)=\frac{1}{\sqrt{2}}(\delta_{\omega,\varepsilon_1}c_1-\delta_{\omega,\varepsilon_3}c_3),\\
   A_3(\omega)&=A_6^\dag(\omega)=\frac{1}{2}(\delta_{\omega,\varepsilon_1}c_1-\sqrt{2}\delta_{\omega,\epsilon_2}c_2+\delta_{\omega,\varepsilon_3}c_3).
\end{align}
Therefore, the time-coarse-grained master equation for this model reads 
\begin{eqnarray}
\nonumber
&&\hskip -.7cm
\frac{d}{d t}\rho_S(t)=-i[\hat{H}_S+\lambda^2\hat{H}^{(\rm LS)}_{\rm tcg}(\Delta t),\rho_S(t)]\\
\nonumber
&&\hskip -.7cm
+\lambda^2\sum_{i,j=1}^3\gamma^+_{i,j}(\Delta t)\left(c_i\rho_S(t)c_j^\dag -\frac{1}{2}\{c_j^\dag c_i,\rho_S(t)\}\right)\\
&&\hskip -.7cm
+\lambda^2\sum_{i,j=1}^3\gamma^-_{i,j}(\Delta t)\left(c_i^\dag\rho_S(t)c_j -\frac{1}{2}\{c_j^\dag c_i,\rho_S(t)\}\right),
\end{eqnarray}
where the Lamb-shift Hamiltonian, $\hat{H}^{(\rm LS)}_{\rm tcg}(\Delta t)$, and $\gamma_{i,j}^{\pm}$ has been define in Eq.~(\ref{eq:lamb3harmonic}) and Eq.~(\ref{eq:gamma}), respectively.

\section{$\Delta t \to \infty$}
\label{App:2}

The results of the GKSL master equation can be recovered by considering the limit where the time interval $\Delta t$ approaches infinity. This is expected because the secular approximation, as derived in the GKSL master equation, effectively averages the time-dependent generator over multiple periods of oscillation. However, this approximation is not necessarily a good one, especially as $\Delta t$ approaches infinity. While the secular approximation simplifies the analysis, it can overlook important dynamical effects, particularly in systems where the interactions with the environment are significant over shorter timescales. In the following, we will prove that the reduced dynamics indeed depends on $\Delta t$; in particular it holds that: 
\begin{equation}
    \lim_{\Delta t\to\infty}\mathcal{D}_{CG}[\bullet]=\mathcal{D}_{\rm GKSL}[\bullet].
\end{equation}
By defining the following function:
\begin{align}
\label{eq:f}
    f(\omega_1,\omega_2)&\equiv\lim_{\Delta t\to \infty}\int_{-\infty}^{+\infty}d\omega\cr
    &\times\operatorname{sinc}\Big((\omega+\omega_1)\frac{\Delta t}{2}\Big)\operatorname{sinc}\Big((\omega+\omega_2)\frac{\Delta t}{2}\Big),\cr
\end{align}
by inserting $\omega + \omega_2 = \omega + \omega_1 + \omega_2 - \omega_1$ to use relations for $\sin(\alpha+\beta)=\sin(\alpha)\cos(\beta)+\cos(\alpha)\sin(\beta)$, Eq.~(\ref{eq:f}) reads
\begin{align}
\label{eq:f2}
    f(\omega_1,\omega_2)&=4\lim_{\Delta t\to \infty}\cos\Big((\omega_2-\omega_1)\frac{\Delta t}{2}\Big)\cr
    &\times\int_{-\infty}^{+\infty}d\omega\frac{\sin^2\Big((\omega+\omega_1)\frac{\Delta t}{2}\Big)}{(\omega+\omega_1)(\omega+\omega_2)\Delta t}\cr
    &+2\lim_{\Delta t\to \infty}\sin\Big((\omega_2-\omega_1)\frac{\Delta t}{2}\Big)\cr
    &\times\int_{-\infty}^{+\infty}d\omega\frac{\sin\Big((\omega+\omega_1)\Delta t\Big)}{(\omega+\omega_1)(\omega+\omega_2)\Delta t}\ .
\end{align}
Then, $f(\omega_1,\omega_2)$ vanishes when $\Delta t\to+\infty$ if $\omega_1\neq\omega_2$. Otherwise,
\begin{align}
\label{eq:f3}
    f(\omega_1)&=\lim_{\Delta t\to \infty}\int_{-\infty}^{+\infty}d\omega\Delta t
    \operatorname{sinc}\Big((\omega+\omega_1)\frac{\Delta t}{2}\Big)\ .
\end{align}
By using the well-known Dirac delta representation
\begin{equation}
\label{Diracdelta}
    \lim_{\Delta t \to \infty}\Delta t \operatorname{sinc}^2\Big((\omega+\omega_1)\frac{\Delta t}{2}\Big)=2\pi\delta(\omega+\omega_1),
\end{equation}
Eq.~(\ref{eq:f}) reads
\begin{equation}
   \label{eq:ffinal} f(\omega_1,\omega_2)=2\pi\delta_{\omega_1,\omega_2}\delta(\omega+\omega_1).
\end{equation}

In the following, by utilizing Eq.~(\ref{eq:ffinal}), the equation presented in Eq.~(\ref{eq:A8b}) in the limit of $\Delta t\to \infty$ can be recast as:
\begin{align}
    \gamma_{k,\ell}^{(\infty)}(\omega_1,\omega_2)&=\lambda^2\sigma_{k\ell}(\omega_1)\delta_{\omega_1,\omega_2},
\end{align}
Therefore, in the limit as $\Delta t$ approaches infinity, the coarse-grained master equation converges to the GKSL master equation as follow:
\begin{equation}
    \frac{\partial}{\partial t}\rho_S(t)=-i[\hat{H}_S,\rho_S(t)]+\mathcal{D}_{\rm GKLS}[\rho(t)],
\end{equation}
where $\mathcal{D}_{\rm GKSL}[\rho_S(t)]$
\begin{eqnarray*}
    \mathcal{D}_{\rm GKSL}[\rho_S(t)]&=&-i[\hat{H}_{LS},\rho_S(t)]\\
    &+&\sum_{\omega,k}\gamma_k(\omega)\Big[A_k(\omega)\rho_S(t)A_k^\dag(\omega)\\
    &-&\frac{1}{2}\{A_k^\dag(\omega) A_k(\omega),\rho_S(t)\}\Big]\ .
\end{eqnarray*}

\section{Continuity equation}
\label{app:Continuity equation}

For a closed three-partite harmonic oscillator chain with Hamiltonian $\hat{H}_S$ as in Eq.~(\ref{eq:HSxp}), its density matrices $\rho_S(t)$ 
evolve according the von Neumann equation $\partial_t\rho_S(t)=-i[\hat{H}_S,\rho_S(t)] $.
Consequently, the time derivative of the probability density $\rho_t(x)={\rm Tr}\big(\hat{n}(x)\rho(t)\big)$ reads
\begin{equation}
    \frac{\partial}{\partial t}\rho_t(x)=\frac{\partial}{\partial t}{\rm Tr}\big(\hat{n}(x)\rho_S(t)\big)=i{\rm Tr}\big([\hat{H}_S,\hat{n}(x)]\rho_S(t)\big),
\end{equation}
with density operator at point $x\in\mathbb{R}$ given by
\begin{equation}
\label{app:prob}
    \hat{n}(x)=\frac{1}{3}\sum_{\ell=1}^3 \delta(x-\hat{x}_\ell)=\frac{1}{6\pi}\sum_{\ell=1}^3\int_{-\infty}^\infty dp_\ell\, e^{i p_\ell(x-\hat{x}_\ell)}\ .
\end{equation}
Since $\displaystyle {\rm e}^{-ip_\ell\hat x_\ell}\hat p_\ell{\rm e}^{ip_\ell\hat x_\ell}=\hat p_\ell+p_\ell$, one computes
\begin{eqnarray*}
\left[\hat{H}_S\,,\,{\rm e}^{-ip_j\hat x_j}\right]&=&-\frac{p_j}{2}\left({\rm e}^{-ip_j\hat x_j}\hat p_j+\hat p_j{\rm e}^{-ip_j\hat x_j}\right)\\
&-&\frac{g}{\omega}p_j\hat p_{2}{\rm e}^{-ip_j\hat x_j}, \quad for \quad j=1,3,\\
\left[\hat{H}_S\,,\,{\rm e}^{-ip_2\hat x_2}\right]&=&-\frac{p_2}{2}\left({\rm e}^{-ip_2\hat x_2}\hat p_2+\hat p_2{\rm e}^{-ip_2\hat x_2}\right)\\
&-&\frac{g}{\omega}p_2\left(\hat p_{1}+\hat p_3\right)\,{\rm e}^{-ip_2\hat x_2}\ ,
\end{eqnarray*}
give rise to
\begin{eqnarray}
&&
i\left[\hat{H}_S\,,\,\hat n(x)\right]=-\partial_x\widehat{j}^{\mathcal{U}}(x),
\label{eq:jUClose}
\end{eqnarray}
where the current operator which comes from the unitary evolution, $\widehat j^{\mathcal{U}}(x)$, reads:
\begin{eqnarray*}
\hskip-.7cm
\widehat{j}^{\mathcal{U}}(x)&=&\frac{1}{6}\sum_{\ell=1}^3\Big(\delta(x-\hat x_\ell)\hat p_\ell+\hat p_\ell\delta(x-\hat x_\ell)\Big)\\
&+&\frac{g}{3\omega}\sum_{\ell=1}^2\Big(\hat p_\ell\delta(x-\hat x_{\ell+1})+\hat p_{\ell+1}\delta(x-\hat x_\ell)\Big)\ .
\end{eqnarray*}
For the open three-harmonic oscillator chain with total Hamiltonian $\hat{H}_T$ as in Eq.~(\ref{Hamtotal}), its density matrices $\rho_T(t)$ 
evolve according the von Neumann equation $\partial_t\rho_T(t)=-i[\hat{H}_T,\rho_T(t)] $.
Consequently, the time derivative of the probability density $\rho_t(x)={\rm Tr}\big(\hat{n}(x)\rho(t)\big)$ reads
$$
    \frac{\partial}{\partial t}\rho_t(x)=i{\rm Tr}\big([\hat{H}_T,\hat{n}(x)]\rho_T(t)\big)\ .
$$
While the commutator of the system Hamiltonian with $\hat{n}(x)$ is the same as Eq.~(\ref{eq:jUClose}), the remaining non-vanishing commutator can be 
computed using 
\begin{eqnarray*}
\left[\hat{H}_{\rm int}\,,\,{\rm e}^{-ip_1\hat x_1}\right]&=&-\frac{p_1}{\sqrt{2\omega_0}}{\rm e}^{-ip_1\hat x_1}\sum_k\gamma_{kL}\big(b_{kL}^\dag-b_{kL}\big),\\
\left[\hat{H}_{\rm int}\,,\,{\rm e}^{-ip_2\hat x_2}\right]&=&0,\\
\left[\hat{H}_{\rm int}\,,\,{\rm e}^{-ip_3\hat x_3}\right]&=&-\frac{p_3}{\sqrt{2\omega_0}}{\rm e}^{-ip_3\hat x_3}\sum_k\gamma_{kR}\big(b_{kR}^\dag-b_{kR}\big),\\
\end{eqnarray*}
that yield $i\lambda\left[\hat{H}_{\rm int}\,,\,\hat n(x)\right]=-\partial_x\widehat{j}^{\mathcal{D}}(x)$, 
where
\begin{align*}
    \widehat{j}^\mathcal{D}(x)&=\frac{\lambda}{3\sqrt{2\omega_0}}\delta(x-\hat{x}_1)\sum_k\gamma_{kL}\big(b_{kL}-b^\dag_{kL}\big)\cr
    &+\frac{\lambda}{3\sqrt{2\omega_0}}\delta(x-\hat{x}_3)\sum_k\gamma_{kR}\big(b_{kR}-b^\dag_{kR}\big).
\end{align*}

In the case of the open harmonic chain described with Eq.~(\ref{eq:CG bosonic chain}), the time-derivative of the probability density is
\begin{equation}
\label{eq:probopen}
    \frac{\partial}{\partial t}\rho_t(x)={\rm Tr}\big(\mathcal{L}^{\rm adj}_{\rm tcg} [\hat{n}(x)]\rho_S(t)\big),
\end{equation}
where $\mathcal{L}^{\rm adj}_{\rm tcg}$ is the dual of the generator in Eq.~(\ref{eq:CG bosonic chain}). The same as the isolated one, we need to see the effect of the generator on the displacement operator,
\begin{align}
\label{C10}
&\mathcal{L}^{\rm adj}_{\rm tcg}[\hat{D}(-i\frac{p_\ell}{\sqrt{2}})]\cr
&=-i\frac{p_\ell}{2\omega_0}\bigg(\omega+S_{\ell\ell}^{(\rm tcg)}(\Delta t)\bigg)\bigg(\hat{p}_\ell\hat{D}(-i\frac{p_\ell}{\sqrt{2}})+\hat{D}(-i\frac{p_\ell}{\sqrt{2}})\hat{p}_\ell\bigg)\cr
&+\frac{p_\ell}{\omega_0}\sum_{k\neq \ell}\bigg(g+i\Re(S_{\ell k}^{(\rm tcg)}(\Delta t))\cr
&+\Im\big(\alpha^{+(\rm tcg)}_{k\ell}(\Delta t)-\alpha^{-(\rm tcg)}_{\ell k}(\Delta t)\big)\bigg)\hat{D}(-i\frac{p_\ell}{\sqrt{2}})\hat{p}_k\cr
&+i\frac{\lambda^2}{4}p_\ell\bigg(\alpha^{+(\rm tcg)}_{\ell\ell}(\Delta t)-\alpha^{-(\rm tcg)}_{\ell\ell}(\Delta t)\bigg)\cr
&\times\bigg(\hat{x}_\ell\hat{D}(-i\frac{p_\ell}{\sqrt{2}})+\hat{D}(-i\frac{p_\ell}{\sqrt{2}})\hat{x}_\ell\bigg)\cr
&+\frac{1}{\omega_0}p_\ell\sum_{k\neq \ell}\bigg(\Im\big(S_{\ell k}^{(\rm tcg)}(\Delta t)\big)\cr
&-i\Re\big(\alpha_{k\ell}^{+(\rm tcg)}(\Delta t)-\alpha_{\ell k}^{-(\rm tcg)}(\Delta t)\big)\bigg)\hat{D}(-i\frac{p_\ell}{\sqrt{2}})\hat{x}_k\cr
&-\frac{1}{4\omega_0}p_\ell^2\bigg(\alpha_{\ell\ell}^{+(\rm tcg)}(\Delta t)+\alpha^{-(\rm tcg)}_{\ell\ell}(\Delta t)\bigg)\hat{D}(-i\frac{p_\ell}{\sqrt{2}})\,,
\end{align}
where $S_{k\ell}^{(\rm tcg)}(\Delta t)$ and $\alpha^{\pm(\rm tcg)}_{\ell k}(\Delta t)$ has been defined in Eqs.~(\ref{eq:SlambCG}--\ref{eq:discoeffCG}).
By using Eqs.~(\ref{op-density}) and (\ref{eq:probopen}--\ref{C10}), the time derivative of the probability density is
\begin{equation}
    \partial_t\rho_t(x)+\partial_x\Big(j^\mathcal{U}_t(x)+j^{\mathcal{D}_{\rm tcg}}_t(x) +\mathcal{Q}^{\rm tcg}_t(x) + \partial_x\, \mathcal{P}^{\rm tcg}_t(x)\Big)=0.
\end{equation}
Here, $j^\mathcal{U}_t(x)$ and $j^{\mathcal{D}_{\rm tcg}}_t(x)$ are derived from the first and second lines of the r.h.s of Equation Eq.~(\ref{C10}), which has been defined in Eqs.~(\ref{prob-current}) and (\ref{eq:jDCG}). Similarly, 
$\mathcal{Q}^{\rm tcg}_t(x)$ and $\mathcal{P}^{\rm tcg}_t(x)$ originate from the third and fourth lines of the r.h.s of Eq.~(\ref{C10}), which has been defined in Eqs.~(\ref{eq:QCG}) and (\ref{eq:PCG}), respectively.

\section{Covariance matrix evolution and time-coarse-graining}
\label{app:CovTcg}

Consider $N$ bosonic modes described by creation and annihilation operators $a^\dagger_i,a_i$ satisfying $[a_i,a^\dag_j]=\delta_{ij}$. Defining the vector $\hat{\xi}=(a_1,...,a_N,a_1^\dag,...,a_N^\dag)^\top$, the commutation relations read $[\xi_i,\xi_j]=\Omega_{ij}$ with symplectic form
\begin{equation}
    \Omega=\begin{pmatrix}\mathbf{0}_N & \mathbb{I}_N \\ -\mathbb{I}_N & \mathbf{0}_N\end{pmatrix}.
\end{equation}
Gaussian states are completely characterized by their first moments $\langle\xi_i\rangle$ and second moments $\langle\xi_i\xi_j\rangle$; these can be obtained by means of the characteristic function $\chi_\rho(Z)=\operatorname{Tr}(e^{-Z^\dagger \mathbb{Z}_{2N} \hat{\xi}}\rho)$, with 
\begin{equation}
    \mathbb{Z}_{2N}=\begin{pmatrix}
      \mathbb{I}_N&\mathbf{0}_N\\
      \mathbf{0}_N&-\mathbb{I}_N
    \end{pmatrix},
\end{equation}
which is Gaussian: $\chi\rho(Z)=e^{-\frac{1}{4}Z^\dag CZ -i d^\top Z}$.
The covariance matrix $C$ and the displacement vector $d$ are given by:
\begin{align}
\label{eq:covdisgen}
C{ij}&=\langle\xi_i\xi_j^\dag+\xi_j^\dag\xi_i\rangle-2\langle\xi_i\rangle\langle \xi_j^\dag\rangle, & d_i&=\langle \xi_i\rangle.
\end{align}
The covariance matrix has the following structure: 
\begin{equation}
\label{eq:covform}
  C=\begin{pmatrix}\mathcal{C}_1 & \mathcal{C}_2 \\ \mathcal{C}_2^\dagger & \mathcal{C}_1^\top\end{pmatrix},  
\end{equation}
 with $\mathcal{C}_1=\mathcal{C}_1^\dagger$, where the off-diagonal elements of each block encode correlations.

Lindblad generators at most quadratic in $a_i,a_i^\dag$ preserve Gaussianity~\cite{OlivaresParis,serafini_quantum_2021} and the generated dynamics is fully defined by the time-evolution of displacement vectors and covariance matrices $d\mapsto d(t)$ and $C\mapsto C(t)$. 
This is the case for bosonic systems evolving in time according to GKSL master equations
\begin{equation}
\label{eq:general masterequation}
    \dot{\rho}=\mathcal{L}[\rho]= \mathcal{L}(\rho)=-i[\hat{H},\rho]+ \lambda^2\mathcal{D}[\rho],
\end{equation}
with $\hat{H}$ at most quadratic in annihilation and creation operators,
\begin{equation}
\label{eq:quadratic-Hamiltonian}
    \hat{H}=\hat{\xi}^\dagger \mathsf{H}\hat{\xi},
\end{equation}
while the dissipative part contains Lindblad operators which are either creation or annihilation:
\begin{equation}
\label{eq:diss}
\mathcal{D}[\bullet]=\sum_{k\ell=1}^{2N}\gamma_{k\ell}(\xi_\ell^\dag\bullet\xi_k-\frac{1}{2}\{\xi_k\xi_\ell^\dag,\bullet\}),
\end{equation}
with positive $\gamma_{k\ell}$.
Using Eqs.~(\ref{eq:covdisgen}), one gets:
\begin{align}
\label{eq:diffcovgen}
    \dot{C}(t)&=\mathcal{M}C(t)+C(t)\mathcal{M}^\dag+\mathcal{N},\\
    \label{eq:diffdisgen}
    \dot{d}(t)&=\mathcal{M}d(t),
\end{align}
with
\begin{align}
\label{eq:MN}
    \mathcal{M}&=i W+\frac{\lambda^2}{2}(\Gamma\mathbb{Z}_{2N}+\mathbb{X}_{2N}\Gamma\Omega),\\
   \mathcal{N}&=\lambda^2(\Gamma-\Omega \Gamma\Omega),\quad [\Gamma]_{ij}=\gamma_{ij},\\
   W&=-\Omega(\mathsf{H}\mathbb{X}_{2N}+\mathbb{X}_{2N}\mathsf{H}),\quad \mathbb{X}_{2N}=\begin{pmatrix}
        {\bf{0}}_N&\mathbb{I}_N\\
        \mathbb{I}_N&{\bf{0}}_N\\
    \end{pmatrix}\ .
\end{align}
A steady state exists and is unique; indeed, time-averaged states under this semigroup are time-invariant~\cite{alicki_quantum_2007}, and only the identity operator commutes with Hamiltonian and Lindblad operators~\cite{spohn_algebraic_1977}. All local, global and time-coarse-grained dynamics are Gaussian, thus ensuring that the unique steady state is Gaussian. 

Moreover, From Eq.~(39), the covariance matrix with the steady state satisfies the following equations:
\begin{equation}
    \mathcal{M} C_\infty + C_\infty \mathcal{M}^\dagger = -\mathcal{N}\, .
\end{equation}
Therefore, in the case of the time-coarse-graining approach, the covariance matrix of the steady state, $C_{\infty}^{(\rm tcg)}$, is the solution of
\begin{equation}
\label{eq:covsteady}
\mathcal{M}_{\rm tcg} C_{\infty}^{(\rm tcg)} + C_{\infty}^{(\rm tcg)} \mathcal{M}^\dagger_{\rm tcg} = -\mathcal{N}_{\rm tcg},
\end{equation}
where the matrices $\mathcal{M}_{\rm tcg}$ and $\mathcal{N}_{\rm tcg}$ are given by
\begin{align}
\label{eq:MNtcg}
    \mathcal{M}_{\rm tcg} &= i W_{\rm tcg} + \frac{\lambda^2}{2}(\Gamma_{\rm tcg} \mathbb{Z}_6 + \mathbb{X}_6 \Gamma_{\rm tcg} \Omega), \\
    \mathcal{N}_{\rm tcg} &= \lambda^2 (\Gamma_{\rm tcg} - \Omega \Gamma_{\rm tcg} \Omega),
\end{align}
with
$$
    W_{\rm tcg} = -\Omega(\mathsf{H}\mathbb{X}_6 + \mathbb{X}_6\mathsf{H}), \
    [\Gamma_{\rm tcg}]_{ij} = (\gamma^-(\Delta t) \oplus \gamma^+(\Delta t))_{ij}\ .
$$
Due to the structure of the problem, the matrices $\mathcal{M}_{\rm tcg}$ and $\mathcal{N}_{\rm tcg}$ are direct sums:
\begin{equation}
\label{eq:MNtcg1}
\mathcal{M}_{\rm tcg} = \mathsf{M}_{\rm tcg} \oplus \mathsf{M}_{\rm tcg}, \quad \mathcal{N}_{\rm tcg} = \mathsf{N}_{\rm tcg} \oplus \mathsf{N}_{\rm tcg}.
\end{equation}
Using Eq.~(\ref{eq:covform}), $\mathcal{M}_{\rm tcg}$ and $\mathcal{M}_{\rm tcg}$ in Eq.~(\ref{eq:MNtcg1}),
the relation in Eq.~(\ref{eq:covsteady}) transforms into two independent matrix equations for the sub-blocks $C_1$ and $C_2$ of $C^{\infty}$:
\begin{align}
\label{eq:c1General}
    \mathsf{M}_{\rm tcg} C_1 + C_1 \mathsf{M}_{\rm tcg}^\dagger &= -\mathsf{N}_{\rm tcg}, \\
\label{eq:c2steadyGeneral}
    \mathsf{M}_{\rm tcg} C_2 + C_2 \mathsf{M}_{\rm tcg} &= \mathbf{0}.
\end{align}
Here, the matrices $\mathsf{M}_{\rm tcg}$ and $\mathsf{M}_{\rm tcg}$ are
\begin{align}
    \mathsf{M}_{\rm tcg} &= -i(H_{\rm d} + \lambda^2 H^{\rm (LS)}_{\rm tcg}) - \frac{\pi}{4}\lambda^2 \mathsf{J}_{\rm tcg},\\
    [\mathsf{N}_{\rm tcg}]_{ij} &= \frac{\pi}{2}\lambda^2(\gamma^-_{ij}(\Delta t) + \gamma^+_{ij}(\Delta t)),
\end{align}
where $H_{\rm d} = \operatorname{diag}(\epsilon_1, \epsilon_2, \epsilon_3)$ and 
\begin{eqnarray*}
[H^{\rm (LS)}_{\rm tcg}]_{ij}&=& S_{ij}^+(\Delta t)+S_{ji}^-(\Delta t)\ , \cr
[\mathsf{J}_{\rm tcg}]_{ij}&=& \gamma^-_{ij}(\Delta t) - \gamma^+_{ij}(\Delta t) \ .
\end{eqnarray*}
Given the uniqueness of the steady state, the solution to Eq.~(\ref{eq:c2steadyGeneral}) is necessarily $C_2 = \mathbf{0}$.
To solve for $C_1$ in Eq.~(\ref{eq:c1General}), we employ the vectorization technique. Rewriting the equation in its vectorized form yields
\begin{equation}
\label{eq:vecsteadyGeneral}
F \ket{C_1} = -\ket{\mathsf{N}}, \quad \text{with} \quad F = \mathsf{M} \otimes \mathbb{I} + \mathbb{I} \otimes \mathsf{M}^*,
\end{equation}
where $\ket{C_1}$ and $\ket{\mathsf{N}}$ are the vectorized forms of $C_1$ and $\mathsf{N}$, respectively. The formal solution is then given by
\begin{equation}
\label{eq:ketc1General}
\ket{C_1} = -F^{-1} \ket{\mathsf{N}}.
\end{equation}
The matrix $C_1$ obtained from this solution fully determines the steady-state properties, including currents and source terms appearing in the continuity equation.

\bibliography{ref}

\end{document}